\documentclass[10pt,twocolumn,english,twocolumn,aps,prd]{revtex4-2}
\usepackage{lmodern}

\usepackage[T1]{fontenc}
\usepackage[latin9]{inputenc}
\usepackage{babel}
\usepackage{float}
\usepackage{amsmath}
\usepackage{graphicx}
\usepackage[unicode=true,
 bookmarks=false,
 breaklinks=false,pdfborder={0 0 1},colorlinks=false,colorlinks=true,
  urlcolor=blue,
  linkcolor=blue,
  citecolor=blue]{hyperref}

\makeatletter
\usepackage{lmodern}
\usepackage[T1]{fontenc}
\usepackage{prettyref}
\usepackage{textgreek}
\usepackage{inputenc}
\usepackage{slashed}
\usepackage{hyperref}
\usepackage{bibentry}
\usepackage{bm}
\usepackage{color}

\makeatother

\begin{document}
\title{Transmutation of protons in a strong electromagnetic field}
\author{T. N. Wistisen}
\author{C. H. Keitel}
\author{A. Di Piazza}
\email{dipiazza@mpi-hd.mpg.de}

\affiliation{Max-Planck-Institut f\"{u}r Kernphysik, Saupfercheckweg 1, Heidelberg
D-69117, Germany}
\begin{abstract}
The process of turning a proton into a neutron, positron
and electron-neutrino in a strong plane-wave electromagnetic field is studied. This
process is forbidden in vacuum and is seen to feature an exponential
suppression factor which is non-perturbative in the field amplitude. The suppression 
is alleviated when the proton experiences a field strength of about ten times the 
Schwinger critical field in its rest frame or larger. Around this threshold the lifetime of the proton, in its rest frame, is comparable to the conventional neutron decay lifetime. As the field strength is increased, the proton lifetime becomes increasingly short. We investigate possible scenarios where
this process may be observed in the laboratory using an ultra-intense
laser and a high-energy proton beam with the conclusion, however, that
it would be very challenging to observe this effect in the near future.
\end{abstract}
\maketitle

\section{Introduction}

In the Standard Model, the proton is regarded as a stable particle
and experimentally it is shown that the lifetime is at least on the order of $10^{33}$
years \cite{MIURA2016516}. The proton is stable in the Standard
Model due to baryon number conservation and to the fact that there is no lighter
baryon to which the proton can decay. The presence of a strong electromagnetic
field, however, allows absorption of four-momentum from the field,
thus allowing the lighter proton to turn into heavier products. Electromagnetic
field strengths on the order of the Schwinger critical field given
by $E_{cr}=m_{e}^{2}c^{3}/e\hbar\approx 1.3\times 10^{16}\;\text{V/cm}$, where $m_{e}$ is the electron
mass, $c$ the speed of light, $e>0$ the elementary charge and $\hbar$
Planck constant, sets the scale at which nonlinear quantum effects in
electrodynamics become important \cite{Ritus,RevModPhys.84.1177}. Among these,
we mention the production of an electron-positron pair by a single photon in a strong 
electromagnetic field \cite{Reiss_1962, Nikishov_1964, Narozhny_2000, Roshchupkin_2001, Reiss_2009, Heinzl_2010b, Mueller_2011b, Titov_2012, Nousch_2012, Krajewska_2013b, Jansen_2013, Augustin_2014, Meuren_2015,PhysRevD.93.085028,PhysRevD.94.013010,PhysRevLett.117.213201,PhysRevLett.108.240406,PhysRevD.82.072002,PhysRevD.98.116002,PhysRevD.97.036021,PhysRevD.98.016005,PhysRevLett.106.020404,Moore1996149,Baier1984231,Kimb83,PhysRevD.101.076017},
or the non-perturbative Schwinger mechanism, where electric fields on the order of or larger than $E_{c}$ will start to spontaneously produce electron-positron
pairs from vacuum \cite{sauter1931verhalten,heisenberg1936folgerungen,Schwinger1951,Brezin_1970,Popov_1971,
Narozhny_2006,Schutzhold_2008,Dunne_2009,Di_Piazza_2009_d,Bulanov_2010_a,PhysRevD.70.053013,PhysRevD.88.045028}.
To be specific we will study the process where a proton turns into
a neutron, a positron, and an electron-neutrino, i.e.,
\begin{equation}
\label{transmutation}
p\rightarrow n+e^{+}+\nu_{e}.
\end{equation}
We will show that this ``proton-transmutation'' process ``turns on'' when the proton experiences an electromagnetic field of about ten times the Schwinger
field strength in its rest frame and that this process features a similar
non-perturbative exponential suppression as the Schwinger mechanism. As we will
elaborate quantitatively below, one can intuitively understand the similar field scale in proton transmutation and in electron-positron pair production as the energy gaps to be overcome are $\sim (m_N+m_e-m_P)c^2\approx 1.8\;\text{MeV}$ and $\sim 2m_ec^2\approx 1\;\text{MeV}$, respectively, with $m_N$, $m_e$, and $m_P$ being the neutron, the electron/positron, and the proton mass. The process has been considered before \cite{ginzburg1965pion,PhysRevLett.87.151301,PhysRevD.56.953,blasone2020beta,PhysRevD.59.094004},
however always with some significant simplifications such as assuming
the particles to be spin-$0$ instead of spin-$\frac{1}{2}$, or using
an interaction like the electromagnetic interaction, preserving parity. Ritus in Ref. \cite{Ritus}, who mainly studied modification of processes already allowed in vacuum, also makes a semi-quantitative estimate of the probability per unit time of proton transmutation by means of analytical continuation in the case of a constant crossed field (see also Ref. \cite{Lyulka_1985}). Finally, for studies about how decay processes due to the weak interaction are influenced by a strong plane wave we refer to the reviews Refs. \cite{Ritus,Akhmedov_2011}.

In this paper, we treat the process using the so-called Vector-Axial Vector (V-A) point
interaction characterized by the Fermi constant $G_F\approx 1.2\times 10^{-5}\;\text{GeV}^{-2}$ \cite{fermi1934versuch,PhysRev.109.1860.2,PhysRev.109.193}
and the particles as spin-$\frac{1}{2}$ point particles in the presence
of a plane-wave field, i.e., we use the Volkov states to describe charged particles \cite{volkov1935class,beresteckij_quantum_2008}. We may use the V-A
point interaction because the energy-momentum transfer in the process is
on the order of the difference between the neutron and the proton mass, which
is much smaller than the masses of the intermediate $W$ boson. Below, we will
also discuss when the approximation of point particle for the proton and
the neutron is acceptable.

The metric tensor $\eta^{\mu\nu}=\text{diag}(+1,-1,-1,-1)$ is used throughout and the Feynman
slash notation indicates the contraction of a four-vector with the Dirac gamma matrices $\gamma^{\mu}$ (the matrix $\gamma^5$ is defined as $\gamma^5=i\gamma^0\gamma^1\gamma^2\gamma^3$) \cite{beresteckij_quantum_2008}. Finally, units with $\hbar=c=1$ are employed.

\section{Theory}

Below, we describe both the proton and the positron by using Volkov states, which are the exact solution of the Dirac equation for a spin-$\frac{1}{2}$ point particle in a plane-wave field \cite{volkov1935class,beresteckij_quantum_2008}. The latter can be described by the four-vector potential $A^{\mu}(\varphi)$ in the Lorenz gauge $\partial_{\mu}A^{\mu}=0$, where $\varphi=kx$, with $k^{\mu}=(\omega,\bm{k})$ being the characteristic wave four-vector ($k^2=0$ and $\omega=|\bm{k}|$) and $x^{\mu}$ the position four-vector. The positron Volkov state wave function is then (for notational simplicity the spin quantum numbers are not explicitly indicated)
\begin{equation}
\psi_{p}(x)=\frac{1}{\sqrt{2\varepsilon_{p}}}\left(1+\frac{\slashed{k}\slashed{\mathcal{A}}(\varphi)}{2kp}\right)v_{p}e^{iS_{p}},\label{eq:volkovstate-1}
\end{equation}
where $p^{\mu}$ is the positron four-momentum quantum number, $\varepsilon_{p}=\sqrt{m_{e}^{2}+\boldsymbol{p}^{2}}$, and $\mathcal{A}(\varphi)=eA(\varphi)$, whereas $S_{p}$ is given by
\begin{equation}
S_{p}=px+\frac{1}{kp}\int^{\varphi}d\varphi'\left(p\mathcal{A}(\varphi')-\frac{1}{2}\mathcal{A}^{2}(\varphi')\right),\label{eq:Sminus-1}
\end{equation}
and $v_p$ is the negative-energy constant bi-spinor \cite{beresteckij_quantum_2008}.

The beta-decay 4-point V-A interaction Hamiltonian is given by \cite{cahn2009experimental,fermi1934versuch,PhysRev.109.1860.2,PhysRev.109.193}
\begin{align}
H_{\text{int}} & =\frac{G_{F}}{\sqrt{2}}\int d^{3}x\bar{\Psi}_{\text{proton}}(x)\gamma^{\mu}(g_{v}+g_{a}\gamma^{5})\Psi_{\text{neutron}}(x)\nonumber \\
 & \times\bar{\Psi}_{\text{electron}}(x)\gamma_{\mu}(1-\gamma^{5})\Psi_{\text{neutrino}}(x)+\text{H.C.},
\end{align}
where each operator $\Psi$ denotes the quantum field which contains the
operators annihilating the particles and creating the anti-particles indicated as indexes and where the numerical parameters $g_v$ and $g_a$ will be set in the numerical computations to the values $g_{v}=1$ and $g_{a}=-1.262$ \cite{cahn2009experimental}. For the proton-transmutation-process in Eq. \eqref{transmutation}, we need the Hermitian conjugate part of $H_{\text{int}}$. Thus, we also need the wave function of the proton in the external plane-wave field, which, assuming the proton asymptotic four-momentum being $P^{\mu}=(\varepsilon_P,\bm{P})=(\sqrt{m_P^2+\bm{P}^2},\bm{P})$, is given by
\begin{equation}
\psi_{P}(x)=\frac{1}{\sqrt{2\varepsilon_{P}}}\left(1+\frac{\slashed{k}\slashed{\mathcal{A}}(\varphi)}{2kP}\right)u_{P}e^{iS_{P}},
\end{equation}
where
\begin{equation}
S_{P}=-Px-\frac{1}{kP}\int^{\varphi}d\varphi'\left(P\mathcal{A}(\varphi')-\frac{1}{2}\mathcal{A}^{2}(\varphi')\right),\label{eq:Sminus-2}
\end{equation}
and where $u_P$ is the positive-energy constant bi-spinor \cite{beresteckij_quantum_2008}. The neutron and the neutrino are neutral and therefore we describe them via the free particle wave functions given by
\begin{equation}
\psi_{Q}(x)=\frac{1}{\sqrt{2\varepsilon_{Q}}}u_{Q}e^{-iQx},
\end{equation}
\begin{equation}
\psi_{q}(x)=\frac{1}{\sqrt{2\varepsilon_{q}}}u_{q}e^{-iqx},
\end{equation}
where $Q^{\mu}=(\varepsilon_Q,\bm{Q})=(\sqrt{m_N^2+\bm{Q}^2},\bm{Q})$ and $q^{\mu}=(\varepsilon_q,\bm{q})=(\sqrt{m_n^2+\bm{q}^2},\bm{q})$ denote the four-momenta of the neutron and the neutrino, respectively (note that we are implicitly assuming the neutrino to be a Dirac-like particle even though later the neutrino mass will be neglected, which is a well-justified approximation accounting for the accuracy of the obtained results, which, for example, do not include the spatial focusing of the laser field). 

Using the Volkov state for the proton implies that we are treating
it as a point particle and this is acceptable as long as the laser field
in the rest frame of the proton has a wavelength significantly longer
than the size of the proton, and that the photon energy is much smaller
than any potential excitation energy of the proton. Of these two requirements,
the latter is the more restrictive one, which corresponds to an energy
of $294$ MeV for the excitation to the delta-baryon.  Assuming a typical value of 1 eV for the laser photon energy, and e.g. a 7 TeV proton as at the Large Hadron Collider (LHC), this translates into roughly a 7 keV photon energy in the rest frame of the proton, significantly smaller than the mentioned model restriction.

Under the above assumptions, the transition matrix element is then
\begin{align}
\mathcal{M}= & -i\frac{G_{F}}{\sqrt{2}}\frac{1}{\sqrt{16\varepsilon_{P}\varepsilon_{Q}\varepsilon_{p}\varepsilon_{q}}}\int d^{4}xe^{i\left(Q+q+p-P\right)x}Y(\varphi)\nonumber \\
 & \times e^{i\int^{\varphi}d\varphi'\left[\frac{p\mathcal{A}(\varphi')}{kp}-\frac{P\mathcal{A}(\varphi')}{kP}+\frac{1}{2}\mathcal{A}^{2}(\varphi')\left(\frac{1}{kP}-\frac{1}{kp}\right)\right]},\label{eq:matrixelement}
\end{align}
where we have defined
\begin{align}
Y(\varphi) & =\bar{u}_{Q}\gamma^{\mu}(g_{v}+g_{a}\gamma^{5})\left(1+\frac{\slashed{k}\slashed{\mathcal{A}}(\varphi)}{2kP}\right)u_{P}\nonumber \\
 & \times\bar{u}_{q}\gamma_{\mu}(1-\gamma^{5})\left(1+\frac{\slashed{k}\slashed{\mathcal{A}}(\varphi)}{2kp}\right)v_{p}.
\end{align}
Now, anticipating that in a plane wave three light-cone momenta are conserved, it is convenient to write this function in terms of its Fourier transform in $\varphi$ and so we define
\begin{align}
\mathcal{Y}(s) & =\frac{1}{2\pi}\int d\varphi Y(\varphi)e^{is\varphi}\nonumber \\
 & \times e^{i\int^{\varphi}d\varphi'\left[\frac{p\mathcal{A}(\varphi')}{kp}-\frac{P\mathcal{A}(\varphi')}{kP}+\frac{1}{2}\mathcal{A}^{2}(\varphi')\left(\frac{1}{kP}-\frac{1}{kp}\right)\right]},
\end{align}
and therefore we can write
\begin{align}
 & Y(\varphi)e^{i\int^{\varphi}d\varphi'\left[\frac{p\mathcal{A}(\varphi')}{kp}-\frac{P\mathcal{A}(\varphi')}{kP}+\frac{1}{2}\mathcal{A}^{2}(\varphi')\left(\frac{1}{kP}-\frac{1}{kp}\right)\right]}\nonumber \\
 & =\int\mathcal{Y}(s)e^{-is\varphi}ds.
\end{align}
By inserting this expression into Eq. (\ref{eq:matrixelement}) and by performing the
integration over $d^{4}x$ we obtain
\begin{align}
\mathcal{M} & =-i\frac{G_{F}}{\sqrt{2}}\frac{(2\pi)^{4}}{\sqrt{16\varepsilon_{P}\varepsilon_{Q}\varepsilon_{p}\varepsilon_{q}}}\nonumber \\
 & \times\int ds\delta^{4}\left(Q+q+p-P-sk\right)\mathcal{Y}(s).
\end{align}
At this point the (spin-resolved) transition probability is given by $dP=|\mathcal{M}|^{2}d^{3}\boldsymbol{Q}d^{3}\boldsymbol{q}d^{3}\boldsymbol{p}/(2\pi)^{9}$. After appropriately taking the square of the delta-function, we obtain the probability as
\begin{align}
dP & =\frac{G_{F}^{2}}{2}\frac{1}{(2\pi)^{4}}\frac{1}{16kP}\int ds\frac{d^{3}\boldsymbol{Q}}{\varepsilon_{Q}}\frac{d^{3}\boldsymbol{q}}{\varepsilon_{q}}\frac{d^{3}\boldsymbol{p}}{\varepsilon_{p}}\nonumber \\
 & \times\delta^{4}\left(Q+q+p-P-sk\right)\left|\mathcal{Y}(s)\right|^{2},\label{eq:probability}
\end{align}
where one may note that each factor is now Lorentz invariant. Now, we turn to the evaluation of the quantity $\left|\mathcal{Y}(s)\right|^{2}$. From
Eq. (\ref{eq:probability}), we have that
\begin{equation}
\left|\mathcal{Y}(s)\right|^{2}=\frac{1}{(2\pi)^{2}}\iint e^{i\Phi(\varphi,\varphi')}Y(\varphi)Y^{\dagger}(\varphi^{\prime})d\varphi d\varphi^{\prime},
\end{equation}
where we defined
\begin{align}
&\Phi(\varphi,\varphi^{\prime})  =s(\varphi-\varphi^{\prime})\nonumber \\
 &+\int_{\varphi^{\prime}}^{\varphi}dx\left[\frac{p\mathcal{A}(x)}{kp}-\frac{P\mathcal{A}(x)}{kP}+\frac{1}{2}\mathcal{A}^{2}(x)\left(\frac{1}{kP}-\frac{1}{kp}\right)\right].
\label{eq:phase}
\end{align}
We therefore have that the probability summed over final spins and
averaged over the proton spin is given by
\begin{align}
dP & =\frac{G_{F}^{2}}{2}\frac{1}{(2\pi)^{6}}\frac{1}{32kP}\int dsd\varphi d\varphi^{\prime}\nonumber \\
 & \times e^{i\Phi(\varphi,\varphi')}\sum_{\text{spins}}Y(\varphi)Y^{\dagger}(\varphi^{\prime})\nonumber \\
 & \times\delta^{4}\left(Q+q+p-P-sk\right)\frac{d^{3}\boldsymbol{Q}}{\varepsilon_{Q}}\frac{d^{3}\boldsymbol{q}}{\varepsilon_{q}}\frac{d^{3}\boldsymbol{p}}{\varepsilon_{p}}.\label{eq:finalprob}
\end{align}
In this first investigation of the process in the presence of a general plane wave we are not interested in polarization effects. Thus, by applying the usual identities for the products of bi-spinors when
summing over spins, we may write [see Appendix (\ref{sec:Appendix-A}) for additional details]
\begin{equation}
\sum_{\text{spins}}Y(\varphi)Y^{\dagger}(\varphi^{\prime})=T^{\mu\nu}(\varphi,\varphi')q^{\alpha}W_{\mu\nu\alpha}(\varphi,\varphi')
\end{equation}
where the tensors $T^{\mu\nu}(\varphi,\varphi')$ and $W_{\mu\nu\alpha}(\varphi,\varphi')$
are given in terms of traces of gamma matrices, and we need only to
keep terms with an even number of gamma matrices, leading to
\begin{widetext}
\begin{align}
T^{\mu\nu}(\varphi,\varphi^{\prime})&=T_{1}^{\mu\nu}(\varphi,\varphi^{\prime})+Q_{\alpha}T_{2}^{\mu\nu\alpha}(\varphi,\varphi^{\prime}),\\
T_{1}^{\mu\nu}(\varphi,\varphi^{\prime})&=m_{N}m_{P}\left(g_{v}^{2}-g_{a}^{2}\right)\text{Tr}\left[\gamma^{\mu}\left(1+\frac{\slashed{k}\slashed{\mathcal{A}}(\varphi)}{2kP}\right)\left(1-\frac{\slashed{k}\slashed{\mathcal{A}}(\varphi^{\prime})}{2kP}\right)\gamma^{\nu}\right],\label{eq:T1}\\
T_{2}^{\mu\nu\alpha}(\varphi,\varphi^{\prime})&=\text{Tr}\left[\gamma^{\alpha}\gamma^{\mu}\left(1+\frac{\slashed{k}\slashed{\mathcal{A}}(\varphi)}{2kP}\right)\slashed{P}\left(1-\frac{\slashed{k}\slashed{\mathcal{A}}(\varphi^{\prime})}{2kP}\right)\gamma^{\nu}\left(g_{v}^{2}+g_{a}^{2}+2g_{v}g_{a}\gamma^{5}\right)\right],\label{eq:T2}\\
W_{\mu\nu\alpha}(\varphi,\varphi')&=2\text{Tr}\left[\gamma_{\alpha}\gamma_{\mu}\left(1+\frac{\slashed{k}\slashed{\mathcal{A}}(\varphi)}{2kp}\right)\slashed{p}\left(1-\frac{\slashed{k}\slashed{\mathcal{A}}(\varphi^{\prime})}{2kp}\right)\gamma_{\nu}(1-\gamma^{5})\right].\label{eq:W}
\end{align}
\end{widetext}
We may now employ the identities from the Appendix (\ref{sec:Appendix-C})
\cite{Ritus} to obtain (for brevity we omit the dependence
on $\varphi$ and $\varphi'$)
\begin{align}
 & \int T^{\mu\nu}q^{\alpha}W_{\mu\nu\alpha}\delta^{4}\left(Q+q+p-P-sk\right)\frac{d^{3}\boldsymbol{q}d^{3}\boldsymbol{Q}}{\varepsilon_{q}\varepsilon_{Q}}\nonumber \\
 & =\frac{J}{2l^{2}}(l^{2}-m_{N}^{2})T_{1}^{\mu\nu}W_{\mu\nu\alpha}l^{\alpha}\nonumber \\
 & +\frac{J}{6l^{4}}\left[l^{2}\left(l^{2}+m_{N}^{2}\right)-2m_{N}^{4}\right]l_{\alpha}l^{\beta}T_{2}^{\mu\nu\alpha}W_{\mu\nu\beta}\nonumber \\
 & +\frac{J}{12l^{2}}\left(l^{2}-m_{N}^{2}\right)^{2}T_{2}^{\mu\nu\alpha}W_{\mu\nu\alpha},\label{eq:TWint}
\end{align}
where (setting for simplicity the neutrino mass to zero)
\begin{align}
J&=\theta(kP-kp)\theta(s-s_{\text{min}})\frac{2\pi\sqrt{\left(l^{2}-m_{N}^{2}\right)^{2}}}{l^{2}},\label{eq:J}\\
l&=sk+P-p,\label{eq:l}\\
s_{\text{min}}&=\frac{m_{N}^{2}-\left(P-p\right)^{2}}{2k\left(P-p\right)},\label{eq:smin}
\end{align}
with $\theta$ denoting the Heaviside function. The expression of $s=s_{\text{min}}$ can be obtained from the kinematical condition $l^{2}>m_{N}^{2}$, which follows from the energy-momentum conservation of the delta-function from Eq. (\ref{eq:finalprob}).
Then, by conveniently setting $s=s_{\text{min}}+\rho$, we obtain
\begin{equation}
l^{2}=2k(P-p)\rho+m_{N}^{2}.\label{eq:lsq}
\end{equation}

The integrals over $\varphi$ and $\varphi'$ in Eq. \eqref{eq:finalprob} can be conveniently turned into a double integral over the central phase  $\varphi_{+}=(\varphi+\varphi')/2$ and over the relative phase $\varphi_{-}=\varphi-\varphi'$. From now on we realistically assume that the plane wave is sufficiently intense that the classical nonlinearity parameter $\xi=eE/m_e\omega\gg 1$ \cite{RevModPhys.84.1177}, where $E$ is the electric-field amplitude of the laser field. It is known that, generally speaking, if $\xi\gg 1$ the largest contribution to the integral in $\varphi_-$ comes from the region $|\varphi_-|\lesssim 1/\xi\ll 1$ \cite{Ritus,RevModPhys.84.1177}. Thus, one can apply the so-called locally constant field approximation (LCFA), where one expands the plane-wave field around $\varphi_-=0$. The LCFA is discussed in detail in the Appendix (\ref{sec:Appendix-B}). We will need to consider both the pre-exponential factor functions and the phase $\Phi(\varphi,\varphi')$ in Eq. \eqref{eq:finalprob} and we start from the latter. Within the LCFA, it is appropriate to expand the phase $\Phi(\varphi,\varphi')$ up to cubic terms in $\varphi_-$ [see Appendix (\ref{sec:Appendix-B})]:
\begin{widetext}
\begin{equation}
\Phi=\tilde{\Phi}+\varphi_{-}\frac{kP}{2k\left(P-p\right)\left(kp\right)}\left(\bm{p}_{\bot}-\frac{kp}{kP}\bm{P}_{\bot}-\frac{k\left(P-p\right)}{kP}\left\langle \bm{\mathcal{A}}_{\perp}\right\rangle \right)^{2},\label{eq:phase-1}
\end{equation}
where
\begin{align}
\left\langle \bm{\mathcal{A}}_{\perp}\right\rangle &=\frac{1}{\varphi-\varphi^{\prime}}\int_{\varphi^{\prime}}^{\varphi}\mathcal{\bm{\mathcal{A}}_{\perp}}(x)dx,\\
\tilde{\Phi}&=\rho\varphi_{-}+\varphi_{-}\frac{(m_{N}^{2}-m_{e}^{2}-m_{P}^{2})(kp)(kP)+m_{e}^{2}(kP)^{2}+m_{P}^{2}(kp)^{2}}{2k\left(P-p\right)(kp)(kP)} +\frac{k(P-p)}{2(kP)(kp)}\left(\frac{d\bm{\mathcal{A}}_{\perp}}{d\varphi_{+}}\right)^{2}\frac{\varphi_{-}^{3}}{12}.\label{eq:reducedphase}
\end{align}
\end{widetext}
Here, we have exploited the additional gauge freedom and standard initial conditions on $A^{\mu}(\varphi)$ to set the time-component and the space-component parallel to $\bm{k}$ of the laser four-vector potential to zero, such that the latter has only non-vanishing transverse components with respect to $\bm{k}$.

At this point, the quantity $T^{\mu\nu}(\varphi,\varphi')q^{\alpha}W_{\mu\nu\alpha}(\varphi,\varphi')$ can be evaluated within the LCFA. The computation of the three terms in Eq. (\ref{eq:TWint}) and the resulting integrals over $\bm{p}_{\bot}$ and $\varphi_{-}$ are straightforward but lengthy and we refer to the Appendix (\ref{sec:Appendix-D}) for details. Here, we mention that these integrals can be taken analytically. Concerning the integration over $d^{2}\bm{p}_{\bot}$, the phase depends on $\bm{p}_{\bot}$ quadratically and the pre-exponential factor contains only powers of $\bm{p}_{\bot}$. Thus, this integration can be carried out by using well-known identities for Gaussian integrals. Now, after integrating over $\bm{p}_{\bot}$, only the reduced phase $\tilde{\Phi}$ remains in the exponent [see Eq. (\ref{eq:reducedphase})], such that we are left with integrals of the form $\int_{-\infty}^{\infty}\varphi_{-}^{n}e^{i(a\varphi_{-}+b\varphi_{-}^{3})}d\varphi_{-}$, which can be expressed in terms of modified Bessel functions $K_{\alpha}(\eta)$ of the second kind \cite{NIST_b_2010}. In particular, one can easily show that
\begin{equation}
\int_{-\infty}^{\infty}\varphi_{-}^{n}e^{i(a\varphi_{-}+b\varphi_{-}^{3})}d\varphi_{-}=c^{n+1}f_{n}(\eta),\label{eq:integral}
\end{equation}
where $c=\sqrt{a/(3b)}$, $\eta=2ac/3$ and where
\begin{equation}
f_{n}(\eta)=\int_{-\infty}^{\infty}z^{n}e^{i\frac{3}{2}\eta(z+\frac{1}{3}z^{3})}dz.\label{eq:f_n}
\end{equation}
In particular we will need
\begin{align}
if_{1}(\eta)&=-\frac{2}{\sqrt{3}}K_{2/3}(\eta),\\
if_{-1}(\eta)&=\frac{2}{\sqrt{3}}\int_{\eta}^{\infty}K_{1/3}(z)dz,\\
f_{-2}(\eta)&=\sqrt{3}\eta\left(\int_{\eta}^{\infty}K_{1/3}(z)dz-K_{2/3}(\eta)\right).
\end{align}
Concerning the convergence of the integrals $f_{-1}(\eta)$ and $f_{-2}(\eta)$, we recall that in taking the Gaussian integrals over $\bm{p}_{\bot}$ one implicitly assumes that the coefficient of $\bm{p}_{\bot}^2$ in the phase has a infinitesimally small positive imaginary part, which then implies that the variable $z$ in the denominators of the integrands in $f_{-1}(\eta)$ and $f_{-2}(\eta)$ has to be intended to be shifted as $z+i0$. The above results show that the process will be exponentially suppressed when $\eta$ is large as $K_{\alpha}(\eta)\sim e^{-\eta}\sqrt{\pi/2\eta}$ for large values of $\eta$, which will correspond to relatively low plane-wave field strengths \cite{NIST_b_2010}. If we consider the process from the rest frame of the proton around the threshold of $\eta\sim 1$, the particles
will be produced as only mildly relativistic, and therefore in the
laboratory frame the produced positron will have an energy of the order of $\gamma_{P}m_{e}$, where $\gamma_{P}$ is the Lorentz factor of the proton. This means that
the natural variable to be introduced to describe the positron is 
\begin{equation}
\zeta=\frac{m_{P}}{m_{e}}\frac{kp}{kP},\label{eq:xvar}
\end{equation}
which will then be of the order of unity near the threshold.

In this way obtain from Eq. \eqref{eq:reducedphase} that
\begin{equation}
\eta =\frac{2}{3}\frac{1}{\chi_{P}}\frac{y^{3}}{\zeta[1-(m_e/m_P)\zeta]^2},\label{eq:etavar}
\end{equation}
where
\begin{align}
\label{eq:yvar}
y&=\sqrt{\frac{l^{2}-m_{e}^{2}-m_{P}^{2}}{m_{e}m_{P}}\zeta+1+\zeta^{2}},\\
\chi_{P}&=\frac{e\sqrt{-(F^{\mu\nu}P_{\nu})^{2}}}{m_{P}m_{e}^{2}}=\frac{(kP)}{m_{P}m_{e}^{2}}\bigg|\frac{d\bm{\mathcal{A}}_{\perp}}{d\varphi_{+}}\bigg|.
\end{align}
Here, $\chi_{P}$ is the ratio of the field strength experienced by
the proton in its rest frame and the Schwinger field strength $E_{cr}$ \cite{Ritus,RevModPhys.84.1177}.

Now, in order to obtain a more compact expression, we keep only the leading-order terms and neglect terms suppressed by the small factors $m_{e}/m_{P}$ and/or $m_{e}/m_{N}$ in the pre-exponent. For the sake of later convenience we keep the exponent exact in these ratios. Also, we assume the plane wave to be linearly polarized. Under these conditions and by introducing the proton proper time $\tau$ via the relation $d\varphi_{+}=(kP/m_{P})d\tau$, we obtain [see Appendix (\ref{sec:Appendix-D}) for additional details]
\begin{widetext}
\begin{align}
\frac{dP}{d\tau d\zeta} & =\frac{G_{F}^{2}}{32\pi^4}m_{e}^{5}m_{P}^{2}\theta(m_P/m_e-\zeta)\int_0^{\infty} dz\frac{z^{2}}{8l^{4}\zeta^{4}}\left[1+\zeta^{2}+\left(\frac{m_{N}^{2}-m_{e}^{2}-m_{P}^{2}}{m_{e}m_{P}}\right)\zeta\right]^{3}\nonumber \\
 & \times\Biggl\{2m_{N}m_{P}\left(g_{v}^{2}-g_{a}^{2}\right)\left[y^{2}if_{1}-\left(1+\zeta^{2}\right)if_{-1}+\zeta\frac{\chi_{P}}{y}f_{-2}\right]\nonumber \\
 & +\frac{1}{3}\left(1+\frac{2m_{N}^{2}}{l^{2}}\right)\left(g_{v}^{2}+g_{a}^{2}\right)\left(2m_{P}^{2}+3l^{2}-m_{N}^{2}\right)\nonumber \\
 & \times\left\{-y^{2}if_{1}-\zeta\frac{\chi_{P}}{y}f_{-2}+if_{-1}\left[ 1+\left(1+\frac{2m_{P}^{2}+l^{2}-m_{N}^{2}}{2m_{P}^{2}+3l^{2}-m_{N}^{2}}\frac{l^{2}-m_{N}^{2}}{m_{P}^{2}}\right)\zeta^{2}\right] \right\}\nonumber \\
 & +\frac{l^{2}-m_{N}^{2}}{3}\left[5\left(g_{v}^{2}+g_{a}^{2}\right)-6g_{v}g_{a}\right]\left[if_{-1}\left(1+\zeta^{2}\right)-y^{2}if_{1}-\zeta\frac{\chi_{P}}{y}f_{-2}\right]\Biggr\}.\label{eq:finalprob-1}
\end{align}
\end{widetext}
We observe here that we changed variable from $\rho$ to $z$ by introducing
\begin{equation}
z=\zeta\frac{2\rho kP[1-(m_e/m_P)\zeta]}{\left(m_{N}^{2}-m_{e}^{2}-m_{P}^{2}\right)\zeta+m_{e}m_{P}(1+\zeta^{2})}
\end{equation}
and therefore eliminating $2\rho k(P-p)$ in $l^{2}$ from Eq. (\ref{eq:lsq}) using this expression, we can express $l^{2}$ in terms of the independent variable $\zeta$ and the integration variable $z$. The probability of proton transmutation per unit of proton proper time in a constant crossed field was also computed in Ref. \cite{Lyulka_1985}. Although the comparison of the analytical expressions of the probability is not straightforward, we have ensured numerically in the regime $\chi_P\ll 1$ that Eq. (\ref{eq:finalprob-1}) is in agreement with Eqs. (3)-(6) in Ref. \cite{Lyulka_1985}.

In order to gain insight into the process, we discuss the regime where $\chi_P\ll 1$. This will also give us the possibility of comparing our results with the corresponding analytical expression of the total probability per unit time obtained in Ref. \cite{Ritus} in this regime (as we have mentioned, we have agreement with the results in Ref. \cite{Lyulka_1985}) by analytically continuing the same quantity of the decay of a charged pion into a neutral pion, an electron (or positron depending on the charge of the initial pion) and an anti-neutrino (a neutrino) in the presence of a constant crossed field. Note that this latter decay does occur also in vacuum as the charged pions are heavier than the neutral pion and an electron/positron. Now, our analysis above Eq. (\ref{eq:xvar}) suggests that the process in this regime is exponentially suppressed as $\exp(-\eta)$, as it can be ascertained from the asymptotic expression of the modified Bessel functions at large values of the argument \cite{NIST_b_2010}. The total probability $dP/d\tau$ per unit time is here expressed as a double integral in $z$ and $\zeta$ and, in order to obtain the asymptotic expression of $dP/d\tau$ at $\chi_P\ll 1$ we first compute the exponent $\eta$ at the values $z^*$ and $\zeta^*$ within the integration region that mostly contribute to the integral. This is easily done in the case of $z$ because, from Eq. (\ref{eq:lsq}), we know that the largest contribution comes from the lowest limit of integration, i.e., from the point $z^*=0$, corresponding to $l^2=m_N^2$. Thus, it is convenient to introduce the quantity [see Eq. (\ref{eq:yvar})]
\begin{equation}
C=\frac{m_N^2-m_e^2-m_P^2}{m_e m_P}.
\end{equation}
In the case of the variable $\zeta$ the procedure is complicated by the non-monotonic dependence of $\eta$ on $\zeta$. In this case the point of maximum contribution is obtained by applying the stationary-phase method, i.e., by solving the equation $\partial\eta/\partial \zeta|_{\zeta=\zeta^*}=0$, where we have already set $l^2=m_N^2$ [see Eq. (\ref{eq:yvar})]. The only positive root of this equation is
\begin{widetext}
\begin{equation}
\label{zeta_star}
\zeta^*=\frac{\sqrt{\left(C+6\frac{m_e}{m_P}\right)^2+32\left(1+\frac{3}{4}\frac{m_e}{m_P}C\right)}-\left(C+6\frac{m_e}{m_P}\right)}{8\left(1+\frac{3}{4}\frac{m_e}{m_P}C\right)}\approx 0.32
\end{equation}
\end{widetext}
and we have proved analytically that by indicating as $\eta^*$ the value
\begin{equation}
\label{eta_star}
\eta^* =\frac{2}{3}\frac{1}{\chi_{P}}\frac{(C\zeta^*+1+\zeta^{*\,2})^{3/2}}{\zeta^*[1-(m_e/m_P)\zeta^*]^2},
\end{equation}
of $\eta$ at $z=z^*$, i.e., $l^2=m_N^2$ and at $\zeta=\zeta^*$, we exactly obtain Ritus' exponential, which is given by the quantity $(2/3)z_0^{3/2}$ in the last equation on page 578 in Ref. \cite{Ritus}, where
\begin{equation}
\begin{split}
z_0&=\frac{3}{4}\left(\frac{m_N}{2m_P}\right)^{2/3}\left(3+\sqrt{1+\frac{8}{\delta^2}}\right)\left(\sqrt{1+\frac{8}{\delta^2}}-1\right)^{1/3}\\
&\quad\times\left(\frac{\delta^2}{\chi_P}\right)^{2/3}
\end{split}
\end{equation}
with $\delta=(m_P^2-m_e^2-m_N^2)/2m_em_N$. We also report the first two terms of the expansion of $\eta^*$ in the parameter $m_e/(m_N-m_P)\approx 0.4$:
\begin{equation}
\begin{split}
\eta^*&\approx \frac{\sqrt{3}}{\chi_P}\frac{m_N^2-m_P^2}{m_em_P}\left[1+\frac{3m_N^2-m_P^2}{(m_N+m_P)^2}\left(\frac{m_e}{m_N-m_P}\right)^2\right]\\
&\approx \frac{\sqrt{12}}{\chi_P}\frac{m_N-m_P}{m_e}\left[1+\frac{1}{2}\left(\frac{m_e}{m_N-m_P}\right)^2\right]\approx \frac{9.5}{\chi_P}.
\end{split}
\end{equation}
This is already a good approximation of the exact result, which, at the same accuracy as the above equation, reads $\eta^*\approx 9.4/\chi_P$. In this respect, we can conclude that the process ``turns on'' at values of $\chi_P$ of about ten, which agrees with Ritus' general estimate of the threshold in Ref. \cite{Ritus}, which in our notation reads $\chi_P\sim \delta_P^2=(m_N-m_P)^2/m_e^2=6.4$. We observe that Ritus arrives to the threshold condition $\chi_P\sim \delta_P^2$ because the quantity $\eta^*$ can also be written as $B\delta_P^2/\chi_P$, with $B$ being a numerical coefficient of the order of unity (in fact, we have checked that it is $B\approx 9.4/6.4\approx 1.5$ in agreement with our results). 

By following the above saddle-point approach, we can also obtain the analytical asymptotics of the pre-exponential function at $\chi_P\ll 1$. This can be achieved by evaluating the pre-exponential functions at the points $z^*=0$ [except, of course, the overall function $z^2$, see Eq. (\ref{eq:finalprob-1})] and $\zeta^*$ and by expanding the exponent $\eta$ first up to first order in $z$ and then up to the second order in $\zeta$ (recall that by definition the first-order term of the expansion in $\zeta$ vanishes at $\zeta=\zeta^*$). The resulting exponential (in $z$) and Gaussian (in $\zeta$) integrals can be easily taken and the asymptotic expression of the probability per unit of proton proper time at $\chi_P\ll 1$ reads:
\begin{widetext}
\begin{equation}
\begin{split}
\frac{dP}{d\tau}&=\frac{3G_F^2}{64\pi^3}m_e^5\left(\frac{m_P}{m_N}\right)^2\frac{(m_P^2+m_N^2)(g_v^2+g_a^2)+m_Nm_P(g_a^2-g_v^2)}{m_N^2}\frac{\zeta^*[C\zeta^*+2(1+\zeta^{*\,2})]}{\sqrt{(C\zeta^*+1-8\zeta^{*\,2})(C\zeta^*+1+\zeta^{*\,2})^5}}\\
&\quad\times\chi_P^4\exp\left[-\frac{2}{3}\frac{1}{\chi_P}\frac{(C\zeta^*+1+\zeta^{*\,2})^{3/2}}{\zeta^*}\right]\approx 5.0\times 10^{-6}\chi_P^4\exp\left(-\frac{9.4}{\chi_P}\right)\text{[s$^{-1}$]}
\end{split}
\end{equation}
\end{widetext}
Note that we have kept the corrections scaling as $m_e/m_P$ in Eqs.  (\ref{zeta_star}) and (\ref{eta_star}) to compare analytically the exponent with the corresponding result by Ritus. Here, we have ignored these corrections because, as we have mentioned, they were already ignored in the pre-exponential function in Eq. (\ref{eq:finalprob-1}). In order to compare the above expression with the corresponding Ritus's result in the last equation of page 578 in Ref. \cite{Ritus} (this analytical asymptotics was not obtained in Ref. \cite{Lyulka_1985}), we observe that that equation can be written in our notation as
\begin{equation}
\begin{split}
\frac{dP_R}{d\tau}&=G_F^2m_P^5\left(\frac{m_N}{m_P}\right)^8\left(\frac{\chi_P}{\delta^2}\right)^4\exp\left(-\frac{2}{3}z_0^{3/2}\right)\\
&\approx 4.6\times 10^{-6}\left(\frac{m_P}{m_e}\right)^5\chi_P^4\exp\left(-\frac{9.4}{\chi_P}\right)\text{[s$^{-1}$]}
\end{split}
\end{equation}
showing a numerical discrepancy of the order of $(m_e/m_P)^5$. Considering the agreement of our results with those in Ref. \cite{Lyulka_1985} and having proved that at the threshold $\chi_P\sim 10$, the total decay probability per unit time is, as expected, of the same order of magnitude of the conventional neutron beta decay, we conclude that an addition factor $(m_e/m_P)^5$ is missing in the result in Ref. \cite{Ritus} (note that Ritus describes his formula as being ``accurate to within a numerical factor'').

Notice that the considerations about the threshold of the process have been carried out only from the analysis of the Bessel functions and their asymptotic exponential behavior. These considerations would be unchanged if we had considered the easier Fermi model of weak interaction \cite{fermi1934versuch}, as it only relies on the phases of the particles' states in the plane wave. However, a quantitatively more accurate evaluation of the probability requires the use of the more realistic V-A theory of weak interaction.

\section{Numerical results and discussion}
\begin{figure}
\includegraphics[width=1\columnwidth]{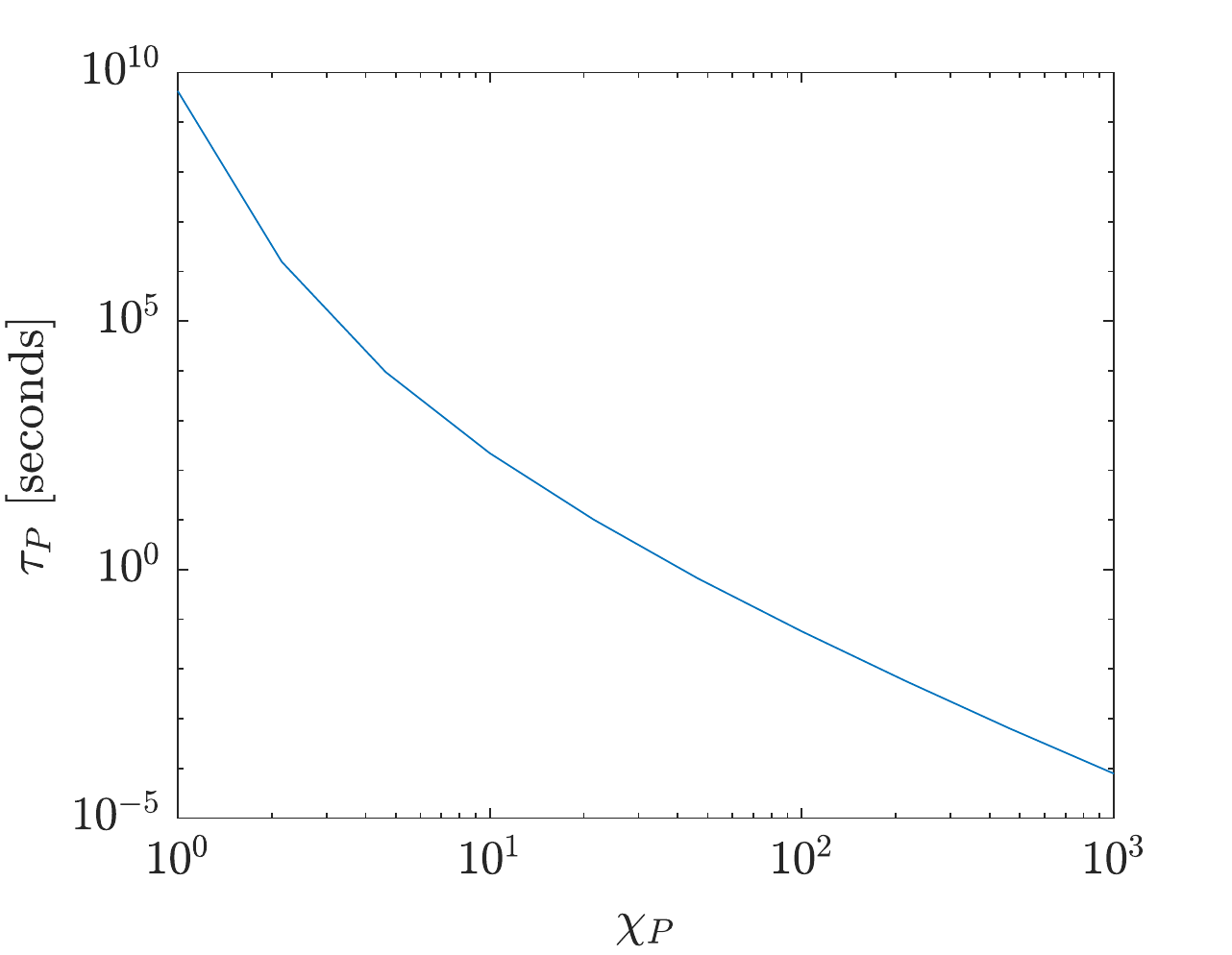}
\caption{The lifetime $\tau_P$ of the proton in its rest frame as a function
of the parameter $\chi_{P}$. \label{fig:The-lifetime-of}}
\end{figure}

Below, we report and discuss the results of numerical evaluation of the proton transmutation formula found above in Eq. (\ref{eq:finalprob-1}). In Fig. (\ref{fig:The-lifetime-of}) we show a plot of the proton lifetime $\tau_P=(\int d\zeta\, dP/d\tau d\zeta)^{-1}$ in the rest frame of the proton. We point out that this result depends solely on $\chi_P$ and therefore that there is no dependence on the laser pulse shape in this figure. The quantity $\tau_P$ in Fig. (\ref{fig:The-lifetime-of}) has to be interpreted as the proton lifetime in a constant crossed field of amplitude $E$. The total proton transmutation probability $P$ in a plane-wave pulse with a given field shape $E(\varphi)$ is obtained by going back to the variable $\varphi_+$ and by taking the double integral $P=\int d\zeta d\varphi_+\, dP/d\varphi_+ d\zeta$, with $\chi_P\to \chi_P(\varphi_+)=2\gamma_{P}|E(\varphi_+)|/E_{cr}$.
As expected, the figure shows a rapid decrease of the proton lifetime for increasing values of $\chi_P$. In particular, the lifetime increases rapidly below a certain threshold. By fitting the lifetime for $10<\chi_P<10^3$, we found that it features a power-law dependence scaling roughly as $\chi_{P}^{-3}$. In Ref. \cite{Lyulka_1985} the asymptotic expression of the lifetime corresponding to a scaling as $1/[\chi_{P}^2\log(\chi_P)]$ with logarithmic accuracy for $\chi_P\gg 1$ is reported. Although, as we will indicate below, our model is not valid for values of $\chi_P\gtrsim 10^3$, in order to compare with that analytical result, we have computed the lifetime for 60 values of $\chi_P$ between $10^3$ and $10^4$. By fitting these values with a function proportional to $1/\{\chi_{P}^a[\log(\chi_P)+b]\}$, with $a$ and $b$ being constant, and we found $a\approx 1.9$ in good agreement with the analytical asymptotic (the constant $b$ was included because in the mentioned range of $\chi_P$ the logarithmic accuracy turned out to be too poor).

Unlike the results in Fig. \ref{fig:The-lifetime-of}, the numerical examples below are obtained for a specific laser pulse shape. We have chosen the Gaussian pulse form given by
\begin{equation}
A^{\mu}(\varphi)=a^{\mu}\text{sin}\left(\varphi\right)e^{-\frac{\varphi^{2}}{2\sigma^{2}}},
\label{a}
\end{equation}
where $a^{\mu}=(0,A,0,0)$, with $A=E/\omega$ and with $\sigma$ describing the pulse duration. Also, we assume a head-on collision between an ultrarelativistic proton and the laser pulse such that $\chi_{P}\approx 2\gamma_{P}E/E_{cr}$.

\begin{figure}
\includegraphics[width=1\columnwidth]{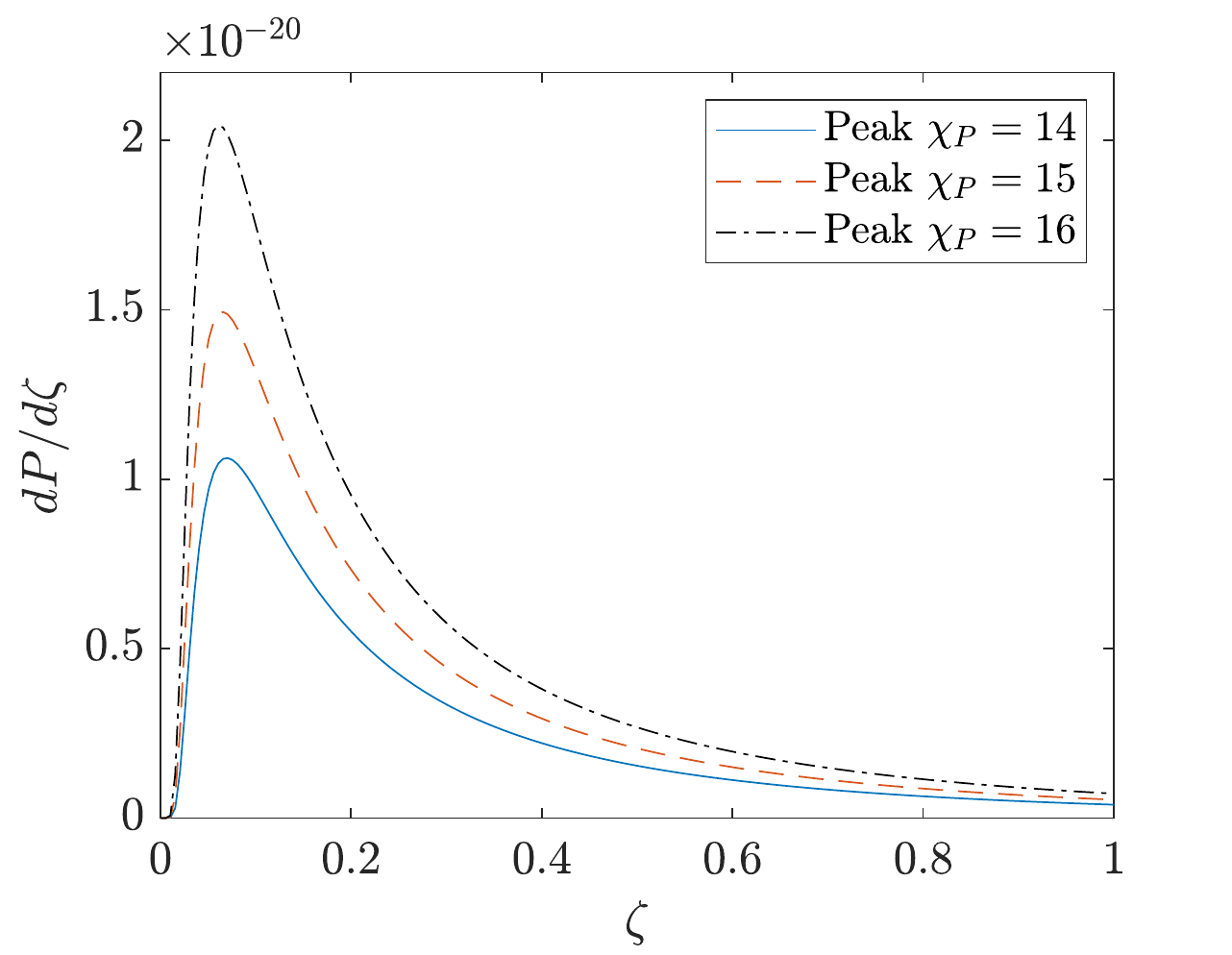}
\caption{The probability spectrum of the emitted positron as a function of $\zeta=m_{P}/m_{e}\times kp/kP$ for a 10-cycle Gaussian pulse [$\sigma=10$ in Eq. (\ref{a})] and for different peak values of $\chi_{P}$. See the text for the remaining numerical parameters. \label{fig:The-probability-spectrum}}
\end{figure}

In Fig. \eqref{fig:The-probability-spectrum} we show examples of the distribution of the positrons for different peak values of the field strength.
For the examples in Fig. \eqref{fig:The-probability-spectrum} we have used $\omega=1.0$ eV,
$\sigma=10$ and $\varepsilon_{P}=7$ TeV. In the cases shown, the spectra show a peak for $\zeta\sim 0.1$, however for larger values of $\chi_P$ we have seen that the peak moves towards lower values of $\zeta$. On the contrary we have ascertained numerically that at values of $\chi_P$ smaller than unity the peak moves towards the point corresponding to $\zeta^*$ as given by Eq. (\ref{zeta_star}). In Table (\ref{tab:Total-probabilities.-Here}) we show the expected results in terms of probability per proton and per collision corresponding to different experimental setups. It is seen that even with future laser facilities brought together with a proton synchrotron such as the LHC, the reaction probability remains small.
\begin{table*}
\begin{tabular}{|c|c|c|c|c|c|}
\hline 
 & HL-LHC standard laser & HL-LHC exawatt laser & HL-LHC exawatt laser & FCC exawatt laser & FCC exawatt laser\tabularnewline
\hline 
\hline 
$\xi$  & 100 & 2350 & 7400 & 2350 & 7400\tabularnewline
\hline 
$I$ {[}W/$\text{cm}^{2}${]} & $1.85\times10^{22}$ & $1.0\times10^{25}$ & $1.0\times10^{26}$ & $1.0\times10^{25}$ & $1.0\times10^{26}$\tabularnewline
\hline 
$\varepsilon_{P}$ {[}TeV{]} & 7 & 7 & 7 & 50 & 50\tabularnewline
\hline 
$\chi_{P}$ (peak) & 2.74 & 32.2 & 203 & 230 & 724\tabularnewline
\hline 
$N_P$ {[}$10^{11}${]} & 2.2 & 2.2 & 2.2 & 1.0 & 1.0\tabularnewline
\hline 
Prob. per proton & $2.2\times10^{-26}$ & $9.9\times10^{-19}$ & $3.8\times10^{-17}$ & $5.8\times10^{-17}$ & $1.4\times10^{-15}$\tabularnewline
\hline 
Prob. per collision & $4.8\times10^{-15}$ & $2.2\times10^{-7}$ & $8.3\times10^{-6}$ & $5.8\times10^{-6}$ & $1.4\times10^{-4}$\tabularnewline
\hline 
\end{tabular}

\caption{Total proton transmutation probabilities for different experimental setups (note that HL-LHC stands for High-Luminosity Large Hadron Collider, whereas FCC stands for Future Circular Collider, see also the recent review \cite{Shiltsev_2021} on present and future colliders). The symbol $I$ denotes the laser peak intensity and the symbol $N_P$ the number of protons in the bunch. It is assumed that all protons pass through the center of the laser pulse, i.e., that the proton transverse area
is smaller than the laser pulse focal area. A Gaussian laser pulse shape such as that in Eq. (\ref{a}) is chosen, with $\sigma=10$.\label{tab:Total-probabilities.-Here}}
\end{table*}
The physical reason for the probability being so low is that around the threshold
where the process is no longer exponentially suppressed, i.e., $\chi_{P}\sim 10$,
the lifetime of the proton becomes comparable to that of a free neutron.
More precisely, the proton lifetime at $\chi_{P}=10$ is about 235 seconds, which is exceedingly large as compared to the duration of a typical laser pulse on the order of femto- or pico-seconds.
Furthermore, in order to reach high field strengths the laser pulse has been chosen to propagate in the opposite direction as the proton. This implies that the duration of the laser pulse in the rest frame of the proton becomes Lorentz contracted by the Lorentz factor of the proton. In conclusion, for the proton transmutation to be sizable, one would need large field strengths for extended periods of time, i.e., extremely large laser energies. We point out that in Table (\ref{tab:Total-probabilities.-Here}), the probability per collision is the probability per proton times the number $N_P$ of protons in the bunch, i.e., it is assumed that the transverse area of the proton bunch is significantly smaller than the laser pulse focal area. 

It is physically interesting to observe the following: The Schwinger field 
strength contains the mass of the electron and is typically
associated with the field strength where production of electron-positron
pairs becomes sizable. Therefore, one may rightfully ask why this process, involving
protons and neutrons as well, also turns on when the proton experiences
a field relatively close to the Schwinger field. This is somewhat a coincidence
due to the mass of the neutron and proton differing by about a MeV, i.e.,
by an amount indeed comparable with twice the electron mass, which corresponds to the energy gap to be overcome in electron-positron pair production.

Suppose that instead of the neutron we had considered producing the neutral delta baryon $\Delta^{0}$ with a mass of $m_{\Delta^0}=1232$ MeV. Although the $\Delta^{0}$ is only about 30\% heavier than the neutron with mass $m_{N}=939.6$ MeV, the implication for the threshold of the corresponding process would be much more significant. Indeed, if we apply the findings above to this case, we have that $\eta^*\approx 2303/\chi_{P}$ in this case. This also implies that applying the results obtained in this paper above $\chi_{P}\approx 10^{3}$, is not meaningful, as the $\Delta^{0}$ may be seen as an excitation of the neutron, and therefore the assumption of point like particles in the wave functions is no longer allowed. In addition the emitted positron would also experience a quantum nonlinearity parameter of the order of $10^3$, and radiative corrections for the positron interacting with the laser field are expected to become significant \cite{ritus1970radiative,PhysRevD.20.1313,PhysRevD.21.1176,ritus1981radiative,Ritus,Fedotov_2017,Podszus_2019,Ilderton_2019,Baumann_2019,Blackburn_2019,
Yakimenko_2019,PhysRevLett.124.044801,Di_Piazza_2020_c}. 

Finally, we have also considered the possibility of colliding a proton beam with an XFEL pulse, whose photon energy is typically much larger than in the case of an optical beam. In the case of an XFEL it would be unrealistic to use the above formulas obtained within the LCFA and the opposite regime $\xi\ll 1$ seems more appropriate. Thus, we considered a kinematic situation in which the laser photon energy is high enough that the process is allowed by the absorption of a single photon. In order to obtain an order of magnitude of the resulting transmutations probability, we expanded the probability in Eq. (\ref{eq:finalprob}) including the leading (quadratic) term in the field and computed the first term in the pre-exponent, corresponding to the second line of Eq. (\ref{eq:TWint}). We have found that in the case of the collision of $10\;\text{keV}$ photons with $7\;\text{TeV}$ protons, the cross-section for the process is on the order of $10^{-7}$ picobarn. Even assuming optimistic conditions where the field strength is such that $\xi=1$ and that the pulse contains about $3\times 10^5$ cycles (corresponding to about $120\;\text{fs}$) yields a probability on the order of $10^{-15}$ for conversion, or roughly $10^{-4}$ per collision for a bunch containing $10^{11}$ protons (all passing through the laser spot).

\section{Conclusion}

In conclusion, we have presented the formula for the decay rate of
a proton into a neutron, a positron, and an electron neutrino in the presence
of a strong plane-wave field (proton transmutation). The full V-A interaction 
has been employed, meaning that the particles are treated as spin-$\frac{1}{2}$ and 
that parity-violating effects have been taken into account. We have seen that the process
turns on when the proton experiences a field value about ten times the Schwinger field 
strength, due to the masses of the neutron and proton differing by about a MeV, i.e., 
by an amount comparable with the electron mass. We have argued that the composite nature 
of the neutron and proton can be neglected as long as the external field does not vary 
too rapidly and for values of the quantum nonlinearity parameter $\chi_{P}$ associated with the proton up to $10^{3}$. We have shown that at $\chi_{P}=10^{3}$ the lifetime of the proton is roughly only $50$ microseconds. However, this is still far longer than any realistic strong laser pulse, keeping in mind that this pulse duration should be achieved in the rest frame of the proton. This explains physically why it is challenging to observe the proton transmutation experimentally. Analogous conclusions have been drawn in the case of a collision of a proton beam with an XFEL. However, we have shown that in the case of a strong optical laser field the proton transmutation probability features a non-perturbative dependence on the elementary charge as well as on the laser field strength, which is typical of tunneling-like processes. If, in the future, it becomes possible to drastically increase the number of protons in particle accelerators or the density of laser photons, this mechanism could in principle be an attractive source of anti-neutrino bursts of short duration comparable to the laser pulse duration. One should keep in mind that if the density of laser photons is increased, the limits discussed above on $\chi_P$ should not be exceeded. But increasing the laser intensity would allow to decrease $\gamma_P$ such that the pulse duration in the proton rest frame would not suffer as large a Lorentz contraction.

\begin{acknowledgments}
The authors acknowledge insightful discussions with Brian Reville.
\end{acknowledgments}

\appendix

\begin{widetext}

\section{Computation of $Y(\varphi)Y^{\dagger}(\varphi^{\prime})$ \label{sec:Appendix-A}}

By using the standard properties of the Dirac gamma functions \cite{beresteckij_quantum_2008}, one can write the quantity $Y(\varphi)Y^{\dagger}(\varphi^{\prime})$ as

\begin{align}
 & Y(\varphi)Y^{\dagger}(\varphi^{\prime})\nonumber \\
 & =\bar{u}_{n}\gamma^{\mu}(g_{v}+g_{a}\gamma^{5})\left(1+\frac{\slashed{k}\slashed{\mathcal{A}}(\varphi)}{2kP}\right)u_{p}\bar{u}_{v}\gamma_{\mu}(1-\gamma^{5})\left(1+\frac{\slashed{k}\slashed{\mathcal{A}}(\varphi)}{2kp}\right)v\nonumber \\
 & \times\bar{u}_{p}\left(1-\frac{\slashed{k}\slashed{\mathcal{A}}(\varphi')}{2kP}\right)\gamma^{\nu}(g_{v}+g_{a}\gamma^{5})u_{n}\bar{v}\left(1-\frac{\slashed{k}\slashed{\mathcal{A}}(\varphi')}{2kp}\right)\gamma_{\nu}(1-\gamma^{5})u_{v}\nonumber \\
 & =\bar{u}_{n}\gamma^{\mu}(g_{v}+g_{a}\gamma^{5})\left(1+\frac{\slashed{k}\slashed{\mathcal{A}}(\varphi)}{2kP}\right)u_{p}\bar{u}_{p}\left(1-\frac{\slashed{k}\slashed{\mathcal{A}}(\varphi')}{2kP}\right)\gamma^{\nu}(g_{v}+g_{a}\gamma^{5})u_{n}\nonumber \\
 & \times\bar{u}_{v}\gamma_{\mu}(1-\gamma^{5})\left(1+\frac{\slashed{k}\slashed{\mathcal{A}}(\varphi)}{2kp}\right)v\bar{v}\left(1-\frac{\slashed{k}\slashed{\mathcal{A}}(\varphi')}{2kp}\right)\gamma_{\nu}(1-\gamma^{5})u_{v}\nonumber \\
 & =\text{Tr}\left[\bar{u}_{n}\gamma^{\mu}(g_{v}+g_{a}\gamma^{5})\left(1+\frac{\slashed{k}\slashed{\mathcal{A}}(\varphi)}{2kP}\right)u_{p}\bar{u}_{p}\left(1-\frac{\slashed{k}\slashed{\mathcal{A}}(\varphi')}{2kP}\right)\gamma^{\nu}(g_{v}+g_{a}\gamma^{5})u_{n}\right]\nonumber \\
 & \times\text{Tr}\left[\bar{u}_{v}\gamma_{\mu}(1-\gamma^{5})\left(1+\frac{\slashed{k}\slashed{\mathcal{A}}(\varphi)}{2kp}\right)v\bar{v}\left(1-\frac{\slashed{k}\slashed{\mathcal{A}}(\varphi')}{2kp}\right)\gamma_{\nu}(1-\gamma^{5})u_{v}\right],
\end{align}
where $u_n$ and $u_p$ ($u_v$ and $v$) are the constant bi-spinors corresponding to the neutron and the proton (neutrino and positron), respectively. By summing over the spin of all involved initial and final particles, we obtain
\begin{align}
 & \sum_{\text{spins}}Y(\varphi)Y^{\dagger}(\varphi^{\prime})\nonumber \\
 & =\text{Tr}\left[\left(\slashed{Q}+m_{N}\right)\gamma^{\mu}(g_{v}+g_{a}\gamma^{5})\left(1+\frac{\slashed{k}\slashed{\mathcal{A}}(\varphi)}{2kP}\right)\left(\slashed{P}+m_{P}\right)\left(1-\frac{\slashed{k}\slashed{\mathcal{A}}(\varphi')}{2kP}\right)\gamma^{\nu}(g_{v}+g_{a}\gamma^{5})\right]\nonumber \\
 & \times\text{Tr}\left[\left(\slashed{q}+m_{\nu}\right)\gamma_{\mu}(1-\gamma^{5})\left(1+\frac{\slashed{k}\slashed{\mathcal{A}}(\varphi)}{2kp}\right)\left(\slashed{p}-m_{e}\right)\left(1-\frac{\slashed{k}\slashed{\mathcal{A}}(\varphi')}{2kp}\right)\gamma_{\nu}(1-\gamma^{5})\right]\nonumber \\
 & =\text{Tr}\left[\left(\slashed{Q}+m_{N}\right)\gamma^{\mu}\left(1+\frac{\slashed{k}\slashed{\mathcal{A}}(\varphi)}{2kP}\right)(g_{v}+g_{a}\gamma^{5})\left(\slashed{P}+m_{P}\right)(g_{v}-g_{a}\gamma^{5})\left(1-\frac{\slashed{k}\slashed{\mathcal{A}}(\varphi')}{2kP}\right)\gamma^{\nu}\right]\nonumber \\
 & \times\text{Tr}\left[\left(\slashed{q}+m_{\nu}\right)\gamma_{\mu}\left(1+\frac{\slashed{k}\slashed{\mathcal{A}}(\varphi)}{2kp}\right)(1-\gamma^{5})\left(\slashed{p}-m_{e}\right)(1+\gamma^{5})\left(1-\frac{\slashed{k}\slashed{\mathcal{A}}(\varphi')}{2kp}\right)\gamma_{\nu}\right]\nonumber \\
 & =\text{Tr}\left[\left(\slashed{Q}+m_{N}\right)\gamma^{\mu}\left(1+\frac{\slashed{k}\slashed{\mathcal{A}}(\varphi)}{2kP}\right)\left\{ m_{P}\left(g_{v}^{2}-g_{a}^{2}\right)+\slashed{P}(g_{v}^{2}+g_{a}^{2}-2g_{v}g_{a}\gamma^{5})\right\} \left(1-\frac{\slashed{k}\slashed{\mathcal{A}}(\varphi')}{2kP}\right)\gamma^{\nu}\right]\nonumber \\
 & \times\text{Tr}\left[\left(\slashed{q}+m_{\nu}\right)\gamma_{\mu}\left(1+\frac{\slashed{k}\slashed{\mathcal{A}}(\varphi)}{2kp}\right)2(1-\gamma^{5})\slashed{p}\left(1-\frac{\slashed{k}\slashed{\mathcal{A}}(\varphi')}{2kp}\right)\gamma_{\nu}\right]\nonumber \\
 & =\text{Tr}\left[\left(\slashed{Q}+m_{N}\right)\gamma^{\mu}\left(1+\frac{\slashed{k}\slashed{\mathcal{A}}(\varphi)}{2kP}\right)\left\{ m_{P}\left(g_{v}^{2}-g_{a}^{2}\right)+\slashed{P}(g_{v}^{2}+g_{a}^{2}-2g_{v}g_{a}\gamma^{5})\right\} \left(1-\frac{\slashed{k}\slashed{\mathcal{A}}(\varphi')}{2kP}\right)\gamma^{\nu}\right]\\
 & \times\text{Tr}\left[\slashed{q}\gamma_{\mu}\left(1+\frac{\slashed{k}\slashed{\mathcal{A}}(\varphi)}{2kp}\right)2(1-\gamma^{5})\slashed{p}\left(1-\frac{\slashed{k}\slashed{\mathcal{A}}(\varphi')}{2kp}\right)\gamma_{\nu}\right]\nonumber,
\end{align}
and we set
\begin{align}
T^{\mu\nu}(\varphi,\varphi') & =\text{Tr}\left[\left(\slashed{Q}+m_{N}\right)\gamma^{\mu}\left(1+\frac{\slashed{k}\slashed{\mathcal{A}}(\varphi)}{2kP}\right)\left[ m_{P}\left(g_{v}^{2}-g_{a}^{2}\right)+\slashed{P}(g_{v}^{2}+g_{a}^{2}-2g_{v}g_{a}\gamma^{5})\right] \left(1-\frac{\slashed{k}\slashed{\mathcal{A}}(\varphi')}{2kP}\right)\gamma^{\nu}\right]\nonumber \\
 & =m_{N}m_{P}\left(g_{v}^{2}-g_{a}^{2}\right)\text{Tr}\left[\gamma^{\mu}\left(1+\frac{\slashed{k}\slashed{\mathcal{A}}(\varphi)}{2kP}\right)\left(1-\frac{\slashed{k}\slashed{\mathcal{A}}(\varphi')}{2kP}\right)\gamma^{\nu}\right]\nonumber \\
 & +\text{Tr}\left[\slashed{Q}\gamma^{\mu}\left(1+\frac{\slashed{k}\slashed{\mathcal{A}}(\varphi)}{2kP}\right)\slashed{P}\left(1-\frac{\slashed{k}\slashed{\mathcal{A}}(\varphi')}{2kP}\right)\gamma^{\nu}\left(g_{v}^{2}+g_{a}^{2}+2g_{v}g_{a}\gamma^{5}\right)\right],\\
W_{\mu\nu}(\varphi,\varphi')&=2\text{Tr}\left[\slashed{q}\gamma_{\mu}\left(1+\frac{\slashed{k}\slashed{\mathcal{A}}(\varphi)}{2kp}\right)\slashed{p}\left(1-\frac{\slashed{k}\slashed{\mathcal{A}}(\varphi')}{2kp}\right)\gamma_{\nu}(1-\gamma^{5})\right].
\end{align}

\section{Integrals over the momenta of the neutral particles \label{sec:Appendix-C}}
Let $l_{1}=(\varepsilon_1,\bm{l}_1)=(\sqrt{m_1^2+\bm{l}_1^2},\bm{l}_1)$ and $l_{2}=(\varepsilon_2,\bm{l}_2)=(\sqrt{m_2^2+\bm{l}_2^2},\bm{l}_1)$ be two four-momenta. Let us consider the three integrals
\begin{align}
J & =\int\delta^{4}(l-l_{1}-l_{2})\frac{d^{3}l_{1}d^{3}l_{2}}{\varepsilon_{1}\varepsilon_{2}},\\
J_{\alpha}&=\int l_{1,\alpha}\delta^{4}(l-l_{1}-l_{2})\frac{d^{3}l_{1}d^{3}l_{2}}{\varepsilon_{1}\varepsilon_{2}},\\
J_{\alpha\beta} & =\int l_{1,\alpha}l_{2,\beta}\delta^{4}(l-l_{1}-l_{2})\frac{d^{3}l_{1}d^{3}l_{2}}{\varepsilon_{1}\varepsilon_{2}},
\end{align}
where $l=(l^0,\bm{l})$ is a four-vector and $j=1,2$. Due to the four-dimensional delta function, in order these for integrals not to vanish, it is required that $l^0>m_1+m_2$ and $l^2>(m_1+m_2)^2$. These conditions are equivalent to the conditions $l^0>0$ and $l^2>(m_1+m_2)^2$ and then also to the conditions $l_->0$ and $l^2>(m_1+m_2)^2$, with $l_-=(nl)$. Here, we have introduced the quantity $n^{\mu}$ (see also the next Appendix)
\begin{equation}
n^{\mu}=\left(1,\boldsymbol{n}\right)
\end{equation}
where $\boldsymbol{n}=\frac{\boldsymbol{k}}{\omega}$ is the unit vector along the propagation direction of the plane wave. 

Since under a proper Lorentz transformation the integrals $J$, $J_{\alpha}$, and $J_{\alpha\beta}$ are a scalar, a four-vector, and a tensor, respectively, they can be computed by first working in the frame where $\bm{l}=\bm{0}$, and it can be shown that (see also Ref. \cite{Ritus})
\begin{align}
J &=\theta(l_-)\theta\left(l^2-(m_1+m_2)^2\right)\frac{2\pi\sqrt{\left(l^{2}-m_{1}^{2}-m_{2}^{2}\right)^{2}-4m_{1}^{2}m_{2}^{2}}}{l^{2}},\label{eq:J-1}\\
J_{\alpha}&=\frac{J}{2l^{2}}l_{\alpha}[l^{2}+(m_{1}^{2}-m_{2}^{2})],\label{eq:Ja-1}\\
J_{\alpha\beta} &=\frac{J}{6l^{4}}l_{\alpha}l_{\beta}\left[l^{2}\left(l^{2}+m_{1}^{2}+m_{2}^{2}\right)-2\left(m_{1}^{2}-m_{2}^{2}\right)^{2}\right]+\frac{J}{12l^{2}}g_{\alpha\beta}\left[\left(l^{2}-m_{1}^{2}-m_{2}^{2}\right)^{2}-4m_{1}^{2}m_{2}^{2}\right].\label{eq:Jab-1}
\end{align}

\section{Computation of $s_{\text{min}}$ and the validity of the LCFA\label{sec:Appendix-B}}

In this Appendix we will find a useful expression for $s_{\text{min}}$
and conveniently manipulate the phase in the probability. We will here need some identities. We define $n^{\mu}$ as the quantity which in the laboratory frame is given by
\begin{equation}
n^{\mu}=\left(1,\boldsymbol{n}\right)
\end{equation}
where $\boldsymbol{n}=\frac{\boldsymbol{k}}{\omega}$ is the unit vector along the propagation direction of the plane wave. Then let $v$ be some arbitrary 4-vector and define
\begin{align}
v_{\parallel}&=\boldsymbol{n}\cdot\boldsymbol{v},\\
v_{+}&=(v_{0}+v_{\parallel})/2,\\
v_{-}&=v_{0}-v_{\parallel}=nv.
\end{align}
Then the identity holds
\begin{equation}
2v_{+}v_{-}-\bm{v}_{\bot}^{2}=\left(v_{0}+v_{\parallel}\right)\left(v_{0}-v_{\parallel}\right)-\bm{v}_{\bot}^{2}=v_{0}^{2}-v_{\parallel}^{2}-\bm{v}_{\bot}^{2}=v^{2},
\end{equation}
such that for the positron four-momentum $p$, we have
\begin{equation}
p_{+}=\frac{m_{e}^{2}+\bm{p}_{\bot}^{2}}{2np}
\end{equation}
and
\begin{equation}
\left(P-p\right)^{2}=2(P_{+}-p_{+})(P_{-}-p_{-})-\left(\bm{P}_{\bot}-\bm{p}_{\bot}\right)^{2}.
\end{equation}
In this way, we can rewrite $s_{\text{min}}$ as
\begin{align}
s_{\text{min}} & =\frac{m_{N}^{2}-\left(P-p\right)^{2}}{2k\left(P-p\right)}\nonumber \\
 & =\frac{m_{N}^{2}}{2k\left(P-p\right)}+\frac{kP}{2k\left(P-p\right)\left(kp\right)}\left(\bm{p}_{\bot}^{2}-2\frac{kp}{kP}\bm{P}_{\bot}\cdot \bm{p}_{\bot}\right)+\frac{m_{e}^{2}}{2kp}-\frac{m_{P}^{2}+\bm{P}_{\bot}^{2}}{2kP}+\frac{\bm{P}_{\bot}^{2}}{2k\left(P-p\right)}.
\end{align}
Now, we will consider the quantity $\Phi/(\varphi-\varphi')-\rho$:
\begin{align}
 & \frac{\Phi}{\varphi-\varphi'}-\rho\nonumber \\
 & =s_{\text{min}}+\frac{1}{\varphi-\varphi'}\int_{\varphi'}^{\varphi}dx\left[\frac{p\mathcal{A}(x)}{kp}-\frac{P\mathcal{A}(x)}{kP}+\frac{1}{2}\mathcal{A}^{2}(x)\left(\frac{1}{kP}-\frac{1}{kp}\right)\right]=\frac{kP}{2k\left(P-p\right)(kp)}\nonumber \\
 & \times\left[\left(\bm{p}_{\bot}-\frac{kp}{kP}\bm{P}_{\bot}-\frac{1}{\varphi-\varphi'}\frac{k\left(P-p\right)}{kP}\int_{\varphi'}^{\varphi}\bm{\mathcal{A}}_{\perp}(x)dx\right)^{2}-\left(\frac{kp}{kP}\bm{P}_{\bot}+\frac{1}{\varphi-\varphi'}\frac{k\left(P-p\right)}{kP}\int_{\varphi'}^{\varphi}\bm{\mathcal{A}}_{\perp}(x)dx\right)^{2}\right]\nonumber \\
 & +\frac{m_{N}^{2}}{2k\left(P-p\right)}+\frac{m_{e}^{2}}{2kp}-\frac{m_{P}^{2}+\bm{P}_{\bot}^{2}}{2kP}+\frac{\bm{P}_{\bot}^{2}}{2k\left(P-p\right)}+\frac{1}{\varphi-\varphi'}\int_{\varphi'}^{\varphi}dx\left[\frac{\bm{P}_{\bot}\bm{\mathcal{A}}_{\perp}(x)}{kP}-\frac{1}{2}\bm{\mathcal{A}}_{\perp}^{2}(x)\left(\frac{1}{kP}-\frac{1}{kp}\right)\right].
\end{align}
Now, we analyze the terms
\begin{align}
 & \frac{1}{\varphi-\varphi'}\int_{\varphi'}^{\varphi}\frac{\bm{P}_{\perp}\bm{\mathcal{A}}_{\perp}(x)}{kP}-\frac{1}{2}\bm{\mathcal{A}}_{\perp}^{2}(x)\left(\frac{1}{kP}-\frac{1}{kp}\right)dx-\frac{kP}{2k\left(P-p\right)\left(kp\right)}\left(\frac{kp}{kP}\bm{P}_{\bot}+\frac{1}{\varphi-\varphi'}\frac{k\left(P-p\right)}{kP}\int_{\varphi'}^{\varphi}\bm{\mathcal{A}}_{\perp}(x)dx\right)^{2}\nonumber \\
 & =\frac{k(P-p)}{2(kP)(kp)}\left[ \frac{1}{\varphi-\varphi'}\int_{\varphi'}^{\varphi}\bm{\mathcal{A}}_{\perp}^{2}(x)dx-\left(\frac{1}{\varphi-\varphi'}\int_{\varphi'}^{\varphi}\bm{\mathcal{A}}_{\perp}(x)dx\right)^{2}\right] -\frac{1}{2k\left(P-p\right)}\frac{kp}{kP}\bm{P}_{\bot}^{2}.
\end{align}
Inserting this expression into the previous equation we obtain
\begin{align}
\Phi/(\varphi-\varphi')-\rho & =\frac{\left(m_{N}^{2}-m_{e}^{2}-m_{P}^{2}\right)(kp)(kP)+m_{e}^{2}(kP)^{2}+m_{P}^{2}(kp)^{2}}{2k\left(P-p\right)(kp)(kP)}\\
 & +\frac{kP}{2k\left(P-p\right)\left(kp\right)}\left(\bm{p}_{\bot}-\frac{kp}{kP}\bm{P}_{\bot}-\frac{1}{\varphi-\varphi'}\frac{k\left(P-p\right)}{kP}\int_{\varphi'}^{\varphi}\bm{\mathcal{A}}_{\perp}(x)dx\right)^{2}\\
 & +\left(\frac{k(P-p)}{2(kP)(kp)}\right)\left[ \frac{1}{\varphi-\varphi'}\int_{\varphi'}^{\varphi}\bm{\mathcal{A}}_{\perp}^{2}(x)dx-\left(\frac{1}{\varphi-\varphi'}\int_{\varphi'}^{\varphi}\bm{\mathcal{A}}_{\perp}(x)dx\right)^{2}\right].
\end{align}
Up to this point the calculation has been exact. As we will integrate
over $\bm{p}_{\bot}$, the line above containing $\bm{p}_{\bot}$ will introduce
$\bm{\mathcal{A}}_{\perp}$ in the front factor as we will perform the
substitution
\begin{equation}
\bm{x}_{\bot}=\bm{p}_{\bot}-\frac{kp}{kP}\bm{P}_{\bot}-\frac{1}{\varphi-\varphi'}\frac{k\left(P-p\right)}{kP}\int_{\varphi'}^{\varphi}\bm{\mathcal{A}}_{\perp}(x)dx,
\end{equation}
and then integrate with respect to $\bm{x}_{\bot}$ instead of $\bm{p}_{\bot}$. For this reason, we only need to apply the LCFA to the other terms in the phase. As we have mentioned in the main text, 
we introduce
\begin{align}
\varphi_{-}&=\varphi-\varphi',\\
\varphi_{+}&=\frac{\varphi+\varphi'}{2},
\end{align}
and expanding the field around $\varphi_{+}$ for small values of $|\varphi_{-}|$:
\begin{align}
\int_{\varphi'}^{\varphi}\bm{\mathcal{A}}_{\perp}(x)dx & \approx\varphi_{-}\bm{\mathcal{A}}_{\perp}(\varphi_{+})+\frac{\varphi_{-}^{3}}{24}\frac{d^{2}\bm{\mathcal{A}}_{\perp}}{d\varphi_{+}^{2}},\\
\left(\frac{1}{\varphi_{-}}\int_{\varphi'}^{\varphi}\bm{\mathcal{A}}_{\perp}(x)dx\right)^{2}&\approx\bm{\mathcal{A}}_{\perp}^{2}(\varphi_{+})+\frac{\varphi_{-}^{2}}{12}\bm{\mathcal{A}}_{\perp}(\varphi_{+})\frac{d^{2}\bm{\mathcal{A}}_{\perp}}{d\varphi_{+}^{2}},\\
\frac{1}{\varphi_{-}}\int_{\varphi'}^{\varphi}\bm{\mathcal{A}}_{\perp}^{2}(x)dx &\approx\bm{\mathcal{A}}_{\perp}^{2}(\varphi_{+})+\left[ \bm{\mathcal{A}}_{\perp}(\varphi_{+})\frac{d^{2}\bm{\mathcal{A}}_{\perp}}{d\varphi_{+}^{2}}+\left(\frac{d\bm{\mathcal{A}}_{\perp}}{d\varphi_{+}}\right)^{2}\right]\frac{\varphi_{-}^{2}}{12}.
\end{align}
Therefore we obtain

\begin{align}
\Phi & \approx\rho\varphi_{-}+\varphi_{-}\frac{\left(m_{N}^{2}-m_{e}^{2}-m_{P}^{2}\right)(kp)(kP)+m_{e}^{2}(kP)^{2}+m_{P}^{2}(kp)^{2}}{2k\left(P-p\right)(kp)(kP)}\nonumber \\
 & +\varphi_{-}\frac{kP}{2k\left(P-p\right)\left(kp\right)}\left(\bm{P}_{\bot}-\frac{kp}{kP}\bm{P}_{\bot}-\frac{1}{\varphi-\varphi'}\frac{k\left(P-p\right)}{kP}\int_{\varphi'}^{\varphi}\bm{\mathcal{A}}_{\perp}(x)dx\right)^{2} +\frac{k(P-p)}{2(kP)(kp)}\left(\frac{d\bm{\mathcal{A}}_{\perp}}{d\varphi_{+}}\right)^{2}\frac{\varphi_{-}^{3}}{12}.
\end{align}

\subsection{Validity of the LCFA}

The condition of validity of the LCFA is that the integral over $\varphi_-$ should be formed over a region where $|\varphi_{-}|$ is much smaller than unity \cite{Ritus}. By using the identity 
\begin{equation}
\rho=\frac{z}{2\zeta kP}\left[\left(m_{N}^{2}-m_{e}^{2}-m_{P}^{2}\right)\zeta+m_{e}m_{P}(1+\zeta^{2})\right],
\end{equation}
we obtain
\begin{equation}
\Phi=\frac{\varphi_{-}}{2kP}\left[(1+z)\left(m_{N}^{2}-m_{e}^{2}-m_{P}^{2}+m_{e}m_{P}(\frac{1}{\zeta}+\zeta)\right)+\frac{1}{\zeta}\frac{m_{P}}{m_{e}}\left(\frac{d\bm{\mathcal{A}}_{\perp}}{d\varphi_{+}}\right)^{2}\frac{\varphi_{-}^{2}}{12}\right].
\end{equation}
Either the first or the second term dominates and the integral is formed over the region where $\Phi\lesssim 1$. If we assume that the first term is dominant, the condition then becomes ($\zeta\sim 1$)
\begin{equation}
|\varphi_{-}|=\frac{2kP}{m_{N}^{2}-m_{e}^{2}-m_{P}^{2}+2m_{e}m_{P}}\ll1.
\end{equation}
The parameter describing the laser field strength is given by the so-called classical nonlinearity parameter \cite{Ritus}
\begin{equation}
\xi=\frac{e|a|}{m_{e}}.
\end{equation}
By assuming that $|d\bm{A}_{\perp}/d\varphi|\sim |\bm{A}_{\perp}(\varphi)|$, we obtain
\begin{equation}
\frac{\chi_{P}}{\xi}\approx\frac{kP}{m_{e}m_{P}},\label{eq:ratio}
\end{equation}
and the above condition becomes
\begin{equation}
\frac{\chi_{P}}{\xi}\frac{1}{\frac{m_{N}^{2}-m_{e}^{2}-m_{P}^{2}}{2m_{e}m_{P}}+1}\ll1,
\end{equation}
or, approximately, $3.5\xi\gg\chi_{P}$.
We consider now the case in which the second term in the phase $\Phi$ dominates, i.e., where
\begin{equation}
1\approx\frac{1}{kP}\frac{m_{P}}{m_{e}}\left(\frac{d\bm{\mathcal{A}}_{\perp}}{d\varphi_{+}}\right)^{2}\frac{\varphi_{-}^{3}}{24}.
\end{equation}
By using Eq. \eqref{eq:ratio}, this may be rewritten as
\begin{equation}
\frac{24\chi_{P}}{\xi^{3}}\approx\varphi_{-}^{3},
\end{equation}
and therefore $\varphi_{-}^{3}\ll1$ translates into $\xi^{3}\gg 24\chi_{P}$.

\section{Analytical integrations \label{sec:Appendix-D}}

In this Appendix we go through the analytical integrations over $\varphi_{-}$
and $\bm{p}_{\perp}$ of the terms from Eq. \eqref{eq:TWint}. For the sake of convenience, we will
split up the three lines into three subsections. First, however, it is useful
to rewrite the expressions from Eq. \eqref{eq:T1} to Eq. \eqref{eq:W}.
First, we see that we may rewrite
\begin{equation}
\left(1+\frac{\slashed{k}\slashed{\mathcal{A}}(\varphi)}{2kP}\right)\slashed{P}\left(1-\frac{\slashed{k}\slashed{\mathcal{A}}(\varphi^{\prime})}{2kP}\right)=\frac{1}{2kP}\left(m_{P}^{2}\slashed{k}+\slashed{\Pi}\slashed{k}\slashed{\Pi}^{\prime}\right).
\end{equation}
Here we have set
\begin{equation}
\Pi=P-\mathcal{A}(\varphi)+\left(\frac{P\mathcal{A}}{kP}-\frac{\mathcal{A}^{2}}{2kP}\right)k,
\end{equation}
and exploited that $\left(\slashed{P}-\slashed{\mathcal{A}}(\varphi)\right)\slashed{k}=\slashed{\Pi}\slashed{k}$.
$\Pi^{\prime}$ denotes the replacement $\varphi\rightarrow\varphi^{\prime}$.
We also need
\begin{equation}
\left(1+\frac{\slashed{k}\slashed{\mathcal{A}}(\varphi)}{2kP}\right)\left(1-\frac{\slashed{k}\slashed{\mathcal{A}}(\varphi^{\prime})}{2kP}\right)=1+\frac{\slashed{k}\left(\slashed{\mathcal{A}}(\varphi)-\slashed{\mathcal{A}}(\varphi^{\prime})\right)}{2kP}.
\end{equation}
Similarly, we define
\begin{equation}
\pi=p-\mathcal{A}(\varphi)+\left(\frac{p\mathcal{A}}{kp}-\frac{\mathcal{A}^{2}}{2kp}\right)k,
\end{equation}
and the same identities with the replacement $P\rightarrow p$ hold
if we also replace $\Pi\rightarrow\pi$. Using these identities and
carrying out the trace we obtain
\begin{align}
T_{1}^{\mu\nu} & =m_{N}m_{P}\left(g_{v}^{2}-g_{a}^{2}\right)\text{Tr}\left[\gamma^{\mu}\left(1+\frac{\slashed{k}\left(\slashed{A}(\varphi)-\slashed{A}(\varphi^{\prime})\right)}{2kP}\right)\gamma^{\nu}\right]\nonumber \\
 & =4m_{N}m_{P}\left(g_{v}^{2}-g_{a}^{2}\right)\left[g^{\mu\nu}+\frac{k^{\mu}\left(\mathcal{A}^{\nu}(\varphi)-\mathcal{A}^{\nu}(\varphi^{\prime})\right)-k^{\nu}\left(\mathcal{A}^{\mu}(\varphi)-\mathcal{A}^{\mu}(\varphi^{\prime})\right)}{2kP}\right].
\end{align}
Now, we write
\begin{equation}
T_{2}^{\mu\nu\alpha}=\left(g_{v}^{2}+g_{a}^{2}\right)T_{a}^{\mu\nu\alpha}+2g_{v}g_{a}T_{b}^{\mu\nu\alpha}
\end{equation}
and then using the above identities we have
\begin{align}
T_{a}^{\mu\nu\alpha}&=\frac{1}{2kP}\left[m_{P}^{2}\text{Tr}\left(\gamma^{\alpha}\gamma^{\mu}\slashed{k}\gamma^{\nu}\right)+\text{Tr}\left(\gamma^{\alpha}\gamma^{\mu}\slashed{\Pi}\slashed{k}\slashed{\Pi}^{\prime}\gamma^{\nu}\right)\right],\\
T_{b}^{\mu\nu\alpha}&=\frac{1}{2kP}\left[m_{P}^{2}\text{Tr}\left(\gamma^{\alpha}\gamma^{\mu}\slashed{k}\gamma^{\nu}\gamma^{5}\right)+\text{Tr}\left(\gamma^{\alpha}\gamma^{\mu}\slashed{\Pi}\slashed{k}\slashed{\Pi}^{\prime}\gamma^{\nu}\gamma^{5}\right)\right].
\end{align}
At this point we can carry out the traces to obtain
\begin{align}
\frac{kP}{2}T_{a}^{\mu\nu\alpha} & =(kP)\left[\left(\Pi^{\prime\mu}+\Pi^{\mu}\right)g^{\alpha\nu}+\left(\Pi^{\nu}+\Pi^{\prime\nu}\right)g^{\alpha\mu}-g^{\mu\nu}\left(\Pi^{\alpha}+\Pi^{\prime\alpha}\right)\right]\nonumber \\
 & +k^{\alpha}\left[\Pi^{\prime\nu}\Pi^{\mu}-\Pi^{\prime\mu}\Pi^{\nu}+\left(\Pi\Pi^{\prime}-m_{P}^{2}\right)g^{\mu\nu}\right]\nonumber \\
 & +\Pi^{\alpha}\left(k^{\nu}\Pi^{\prime\mu}-k^{\mu}\Pi^{\prime\nu}\right)+\Pi^{\prime\alpha}\left(k^{\mu}\Pi^{\nu}-k^{\nu}\Pi^{\mu}\right)\nonumber \\
 & -\left(k^{\nu}g^{\alpha\mu}+k^{\mu}g^{\alpha\nu}\right)\left(\Pi\Pi^{\prime}-m_{P}^{2}\right).
\end{align}
For the traces involving $\gamma^{5}=i\gamma^0\gamma^1\gamma^2\gamma^3$ we will need
\begin{equation}
\text{Tr}\left(\gamma^{\mu}\gamma^{\nu}\gamma^{\rho}\gamma^{\sigma}\gamma^{5}\right)=-4i\epsilon^{\mu\nu\rho\sigma},
\end{equation}
where $\epsilon^{\mu\nu\rho\sigma}$ is the Levi-Civita tensor with convention $\epsilon^{0123}=+1$. Then, by using the identity
\begin{equation}
\gamma^{\mu}\gamma^{\nu}\gamma^{\rho}=\gamma^{\mu}g^{\nu\rho}-\gamma^{\nu}g^{\mu\rho}+\gamma^{\rho}g^{\mu\nu}+i\epsilon^{\sigma\mu\nu\rho}\gamma^{5}\gamma_{\sigma},
\end{equation}
we obtain
\begin{align}
\frac{kP}{2}T_{b}^{\mu\nu\alpha} & =m_{P}^{2}\frac{1}{4}\text{Tr}\left(\gamma^{\alpha}\gamma^{\mu}\slashed{k}\gamma^{\nu}\gamma^{5}\right)+\frac{1}{4}\text{Tr}\left(\gamma^{\alpha}\gamma^{\mu}\slashed{\Pi}\slashed{k}\slashed{\Pi}^{\prime}\gamma^{\nu}\gamma^{5}\right)\nonumber \\
 & =-m_{P}^{2}i\epsilon^{\alpha\mu\beta\nu}k_{\beta}-\frac{1}{4}\text{Tr}\left[\left(\gamma^{\nu}g^{\mu\alpha}+\gamma^{\mu}g^{\nu\alpha}-\gamma^{\alpha}g^{\mu\nu}\right)\slashed{\Pi}\slashed{k}\slashed{\Pi}^{\prime}\gamma^{5}\right]-\frac{1}{4}\text{Tr}\left(i\epsilon^{\kappa\nu\alpha\mu}\gamma_{\kappa}\slashed{\Pi}\slashed{k}\slashed{\Pi}^{\prime}\right)\nonumber \\
 & =i\left(g^{\mu\alpha}\epsilon^{\nu\beta\rho\sigma}+g^{\nu\alpha}\epsilon^{\mu\beta\rho\sigma}-g^{\mu\nu}\epsilon^{\alpha\beta\rho\sigma}\right)\Pi_{\beta}k_{\rho}\Pi_{\sigma}^{\prime}+i\epsilon^{\mu\nu\alpha\beta}\left[m_{P}^{2}k_{\beta}+(kP)\left(\Pi_{\beta}+\Pi_{\beta}^{\prime}\right)-\left(\Pi\Pi^{\prime}\right)k_{\beta}\right].
\end{align}
At this point it is useful to write down the symmetric and anti-symmetric
parts of $T_{2}^{\mu\nu\alpha}$ and $W_{\mu\nu\alpha}$, with respect
to the indexes $\mu$ and $\nu$ as a symmetric tensor contracted with an anti-symmetric
tensor vanishes. We obtain
\begin{align}
T_{2,S}^{\mu\nu\alpha} & =g^{\mu\nu}\left\{2\left(g_{v}^{2}+g_{a}^{2}\right)\left[ \frac{k^{\alpha}}{kP}\left(\Pi\Pi^{\prime}-m_{P}^{2}\right)-\left(\Pi^{\alpha}+\Pi^{\prime\alpha}\right)\right] -\frac{4g_{v}g_{a}}{kP}i\epsilon^{\alpha\beta\rho\sigma}\Pi_{\beta}k_{\rho}\Pi_{\sigma}^{\prime}\right\}\nonumber \\
 & +g^{\alpha\nu}\left\{2\left(g_{v}^{2}+g_{a}^{2}\right)\left[\left(\Pi^{\prime\mu}+\Pi^{\mu}\right)-k^{\mu}\frac{\Pi\Pi^{\prime}-m_{P}^{2}}{kP}\right]+\frac{4g_{v}g_{a}}{kP}i\epsilon^{\mu\beta\rho\sigma}\Pi_{\beta}k_{\rho}\Pi_{\sigma}^{\prime}\right\}\nonumber \\
 & +g^{\alpha\mu}\left\{2\left(g_{v}^{2}+g_{a}^{2}\right)\left[\left(\Pi^{\nu}+\Pi^{\prime\nu}\right)-k^{\nu}\frac{\Pi\Pi^{\prime}-m_{P}^{2}}{kP}\right]+\frac{4g_{v}g_{a}}{kP}i\epsilon^{\nu\beta\rho\sigma}\Pi_{\beta}k_{\rho}\Pi_{\sigma}^{\prime}\right\},\\
T_{2,A}^{\mu\nu\alpha} & =\frac{2\left(g_{v}^{2}+g_{a}^{2}\right)}{kP}\left[k^{\alpha}\left(\Pi^{\prime\nu}\Pi^{\mu}-\Pi^{\prime\mu}\Pi^{\nu}\right)+\Pi^{\alpha}\left(k^{\nu}\Pi^{\prime\mu}-k^{\mu}\Pi^{\prime\nu}\right)+\Pi^{\prime\alpha}\left(k^{\mu}\Pi^{\nu}-k^{\nu}\Pi^{\mu}\right)\right]\nonumber \\
 & +\frac{4g_{v}g_{a}}{kP}i\epsilon^{\mu\nu\alpha\beta}\left[(kP)\left(\Pi_{\beta}+\Pi_{\beta}^{\prime}\right)+\left(m_{P}^{2}-\Pi\Pi^{\prime}\right)k_{\beta}\right].
\end{align}
We may obtain the same for $W_{\mu\nu\alpha}$ by setting $g_{v}=1$
and $g_{a}=-1$ and replace $P\rightarrow p$ and $Q\rightarrow q$:
\begin{align}
W_{\mu\nu\tau}^{S} & =g_{\mu\nu}\left\{4\left[ \frac{k_{\tau}}{kp}\left(\pi\pi^{\prime}-m_{e}^{2}\right)-\left(\pi_{\tau}+\pi_{\tau}^{\prime}\right)\right] +\frac{4}{kp}i\epsilon_{\tau\beta\rho\sigma}\pi^{\beta}k^{\rho}\pi^{\prime\sigma}\right\}\nonumber \\
 & +g_{\tau\nu}\left\{4\left[\left(\pi_{\mu}^{\prime}+\pi_{\mu}\right)-k_{\mu}\frac{\pi\pi^{\prime}-m_{e}^{2}}{kp}\right]-\frac{4}{kp}i\epsilon_{\mu\beta\rho\sigma}\pi^{\beta}k^{\rho}\pi^{\prime\sigma}\right\}\nonumber \\
 & +g_{\tau\mu}\left\{4\left[\left(\pi_{\nu}+\pi_{\nu}^{\prime}\right)-k_{\nu}\frac{\pi\pi^{\prime}-m_{e}^{2}}{kp}\right]-\frac{4}{kp}i\epsilon_{\nu\beta\rho\sigma}\pi^{\beta}k^{\rho}\pi^{\prime\sigma}\right\},\\
W_{\mu\nu\tau}^{A} & =\frac{4}{kp}\left[k_{\tau}\left(\pi_{\nu}^{\prime}\pi_{\mu}-\pi_{\mu}^{\prime}\pi_{\nu}\right)+\pi_{\tau}\left(k_{\nu}\pi_{\mu}^{\prime}-k_{\mu}\pi_{\nu}^{\prime}\right)+\pi_{\tau}^{\prime}\left(k_{\mu}\pi_{\nu}-k_{\nu}\pi_{\mu}\right)\right]\nonumber \\
 & -\frac{4}{kp}i\epsilon_{\mu\nu\tau\beta}\left[(kp)\left(\pi^{\beta}+\pi^{\prime\beta}\right)+\left(m_{e}^{2}-\pi\pi^{\prime}\right)k^{\beta}\right].
\end{align}

\subsection{Line 1}

Here, we wish to find
\begin{equation}
\int T_{1}^{\mu\nu}(\varphi,\varphi^{\prime})W_{\mu\nu\alpha}(\varphi,\varphi^{\prime})l^{\alpha}e^{i\Phi}d\varphi_{-}d^{2}\bm{p}_{\perp}=\int T_{1}^{\mu\nu}(\varphi,\varphi^{\prime})W_{\mu\nu\alpha}(\varphi,\varphi^{\prime})l^{\alpha}e^{i\left(\tilde{\Phi}+g\varphi_{-}x_{\bot}^{2}\right)}d\varphi_{-}d^{2}\bm{x}_{\perp}
\end{equation}
where we introduced
\begin{align}
\bm{x}_{\bot}&=\bm{p}_{\bot}-\frac{kp}{kP}\bm{P}_{\bot}-\frac{k\left(P-p\right)}{kP}\left\langle \bm{\mathcal{A}}_{\perp}\right\rangle ,\label{eq:xortho}\\
g&=\frac{kP}{2k\left(P-p\right)\left(kp\right)}.\label{eq:g}
\end{align}
By using the expression from the previous appendix we find
\begin{align}
\frac{T_{1}^{\mu\nu}W_{\mu\nu\alpha}l^{\alpha}}{4m_{N}m_{P}\left(g_{v}^{2}-g_{a}^{2}\right)} & =8\left[\frac{lk}{kp}\left(\pi\pi^{\prime}-m_{e}^{2}\right)-l\left(\pi+\pi^{\prime}\right)-\frac{lk}{kP}\frac{(\bm{\mathcal{A}}_{\perp}-\bm{\mathcal{A}}_{\perp}^{\prime})^{2}}{2}\right]+8\left(\frac{1}{kp}-\frac{1}{kP}\right)i\epsilon_{\alpha\beta\rho\sigma}l^{\alpha}\pi^{\beta}k^{\rho}\pi^{\prime\sigma},\label{eq:Term1}
\end{align}
where we employed identity
\begin{align}
\epsilon_{\mu\nu\alpha\beta}l^{\alpha}\left(\pi^{\beta}+\pi^{\prime\beta}\right)k^{\mu}\left(\mathcal{A}^{\nu}-\mathcal{A}^{\prime\nu}\right) & =\epsilon_{\mu\nu\alpha\beta}l^{\alpha}\left(\pi^{\beta}+\pi^{\prime\beta}\right)k^{\mu}\left(\pi^{\prime\nu}-\pi^{\nu}\right) =2\epsilon_{\mu\nu\alpha\beta}k^{\mu}\pi^{\prime\nu}l^{\alpha}\pi^{\beta},
\end{align}
and the fact that the Levi-Civita symbol contracted with the same vector twice
vanishes. At this point, we will expand in with respect to $\varphi_{-}$ to enforce the LCFA.
Each term from Eq. \eqref{eq:Term1} requires special attention,
however the calculation may also be reused later. We have
\begin{equation}
\pi\pi^{\prime}-m_{e}^{2}=\Pi\Pi^{\prime}-m_{P}^{2}=\frac{1}{2}(\bm{\mathcal{A}}_{\perp}-\bm{\mathcal{A}}_{\perp}^{\prime})^{2}\approx\frac{\varphi_{-}^{2}}{2}\left(\frac{d\bm{\mathcal{A}}_{\perp}}{d\varphi_{+}}\right)^{2},
\end{equation}
and therefore
\begin{align}
\int\left(\pi\pi^{\prime}-m_{e}^{2}\right)e^{i\left(\tilde{\Phi}+g\varphi_{-}\bm{x}_{\bot}^{2}\right)}d\varphi_{-}d^{2}\bm{x}_{\perp} & =\int\frac{\varphi_{-}^{2}}{2}\left(\frac{d\bm{\mathcal{A}}_{\perp}}{d\varphi_{+}}\right)^{2}e^{i\left(\tilde{\Phi}+g\varphi_{-}\bm{x}_{\bot}^{2}\right)}d\varphi_{-}d^{2}\bm{x}_{\perp}\nonumber \\
 & =\int\frac{i\pi}{g}\frac{\varphi_{-}}{2}\left(\frac{d\bm{\mathcal{A}}_{\perp}}{d\varphi_{+}}\right)^{2}e^{i\tilde{\Phi}}d\varphi_{-}\nonumber \\
 & =\frac{i\pi}{g}c^{2}\frac{f_{1}}{2}\left(\frac{d\bm{\mathcal{A}}_{\perp}}{d\varphi_{+}}\right)^{2}.
\end{align}
We remind that 
\begin{align}
c & =\sqrt{a/(3b)}=\sqrt{\frac{\rho+\frac{\left(m_{N}^{2}-m_{e}^{2}-m_{P}^{2}\right)(kp)(kP)+m_{e}^{2}(kP)^{2}+m_{P}^{2}(kp)^{2}}{2k\left(P-p\right)(kp)(kP)}}{3\frac{k(P-p)}{(kP)(kp)}\left(\frac{d\bm{\mathcal{A}}_{\perp}}{d\varphi_{+}}\right)^{2}\frac{1}{24}}}\approx\frac{2}{\left|\frac{d\bm{\mathcal{A}}_{\perp}}{d\varphi_{+}}\right|}\sqrt{2\rho kp+m_{e}^{2}\left(\frac{m_{N}^{2}-m_{e}^{2}-m_{P}^{2}}{m_{e}m_{P}}\zeta+1+\zeta^{2}\right)}\nonumber \\
 & =\frac{2m_{e}y}{\left|\frac{d\bm{\mathcal{A}}_{\perp}}{d\varphi_{+}}\right|}.\label{eq:c}
\end{align}
Where we set $k(P-p)\approx kP$. At this level of approximation we also have that
\begin{equation}
\frac{i\pi}{g}c^{2}\frac{f_{1}}{2}\left(\frac{d\bm{\mathcal{A}}_{\perp}}{d\varphi_{+}}\right)^{2}=2\pi(kp)(2m_{e}y)^{2}\frac{if_{1}}{2}.
\end{equation}
This procedure of integrating over $d^{2}\bm{p}_{\perp}$, expanding $\bm{\mathcal{A}}_{\perp}$
in $\varphi_{-}$, and integrating over $\varphi_{-}$ must be carried out for all the terms.

\subsubsection{The term $l\left(\pi+\pi^{\prime}\right)$}

We have that
\begin{align}
l\left(\pi+\pi^{\prime}\right) & =(P-p+sk)\left(2p-(\mathcal{A}+\mathcal{A}^{\prime})+\frac{1}{kp}\left[p(\mathcal{A}+\mathcal{A}^{\prime})-\frac{\mathcal{A}^{2}+\mathcal{A}^{\prime2}}{2}\right]k\right)\nonumber \\
 & =2Pp\frac{kP}{k(P-p)}-2m_{e}^{2}+(\bm{P}_{\perp}-\bm{p}_{\perp})(\bm{\mathcal{A}}_{\perp}+\bm{\mathcal{A}}^{\prime}{}_{\perp})+\frac{k(P-p)}{kp}\left[\frac{\bm{\mathcal{A}}_{\perp}^{2}+\bm{\mathcal{A}}_{\perp}^{\prime2}}{2}-\bm{p}_{\perp}(\bm{\mathcal{A}}_{\perp}+\bm{\mathcal{A}}^{\prime}_{\perp})\right]\nonumber \\
 & +2\rho kp+kp\frac{m_{N}^{2}-m_{P}^{2}-m_{e}^{2}}{k\left(P-p\right)}.
\end{align}
where we used the equality $s_{\text{min}}=\frac{m_{N}^{2}-\left(P-p\right)^{2}}{2k\left(P-p\right)}$.
Now, use that
\begin{align}
Pp & =P_{+}p_{-}+P_{-}p_{+}-\bm{P}_{\perp}\bm{p}_{\perp}\nonumber \\
 & =\left(kp\frac{m_{P}^{2}+\bm{P}_{\perp}^{2}}{2kP}+kP\frac{m_{e}^{2}+\bm{p}_{\perp}^{2}}{2kp}-\bm{P}_{\perp}\bm{p}_{\perp}\right),
\end{align}
and so we obtain
\begin{align}
l\left(\pi+\pi^{\prime}\right) & =\left(kp\frac{m_{P}^{2}+\bm{P}_{\perp}^{2}}{kP}+kP\frac{m_{e}^{2}+\bm{p}_{\perp}^{2}}{kp}-\bm{P}_{\perp}\bm{p}_{\perp}\right)\frac{kP}{k(P-p)}-2m_{e}^{2}+(\bm{P}_{\perp}-\bm{p}_{\perp})(\bm{\mathcal{A}}_{\perp}+\bm{\mathcal{A}}^{\prime}_{\perp})\nonumber \\
 & +\frac{k(P-p)}{kp}\left[\frac{\bm{\mathcal{A}}_{\perp}^{2}+\bm{\mathcal{A}}_{\perp}^{\prime2}}{2}-\bm{p}_{\perp}(\bm{\mathcal{A}}_{\perp}+\bm{\mathcal{A}}^{\prime}_{\perp})\right]+2\rho kp+kp\frac{m_{N}^{2}-m_{P}^{2}-m_{e}^{2}}{k\left(P-p\right)}.
\end{align}
Now, we replace $\bm{p}_{\perp}=\bm{x}_{\perp}+\frac{kp}{kP}\bm{P}_{\perp}+\frac{k(P-p)}{kP}\left\langle \bm{\mathcal{A}}_\perp\right\rangle $
to obtain
\begin{align}
l\left(\pi+\pi^{\prime}\right) & =\left(kp\frac{m_{P}^{2}}{kP}+kP\frac{m_{e}^{2}}{kp}\right)\frac{kP}{k(P-p)}-2m_{e}^{2}\nonumber \\
 & +\frac{k(P-p)}{kp}\left[\left(\left\langle \bm{\mathcal{A}}_\perp\right\rangle -\frac{\bm{\mathcal{A}}_{\perp}+\bm{\mathcal{A}}^{\prime}_{\perp}}{2}\right)^{2}+\frac{(\bm{\mathcal{A}}_{\perp}-\bm{\mathcal{A}}^{\prime}_{\perp})^{2}}{4}\right]+2\rho kp+kp\frac{m_{N}^{2}-m_{P}^{2}-m_{e}^{2}}{k\left(P-p\right)}\nonumber \\
 & +2\frac{kP}{kp}\bm{x}_{\perp}\left(\left\langle \bm{\mathcal{A}}_\perp\right\rangle -\frac{\bm{\mathcal{A}}_{\perp}+\bm{\mathcal{A}}^{\prime}_{\perp}}{2}\right)+\bm{x}_{\perp}^{2}\frac{(kP)^{2}}{k(P-p)(kp)}.
\end{align}
Once again approximating $k(P-p)\approx kP$ and by neglecting the difference $\left\langle \bm{\mathcal{A}}_{\perp}\right\rangle -\left(\bm{\mathcal{A}}_{\perp}+\bm{\mathcal{A}}^{\prime}{}_{\perp}\right)/2$, we obtain
\begin{align}
l\left(\pi+\pi^{\prime}\right) & \approx\frac{m_{e}m_{P}}{\zeta}\left(1+\zeta^{2}\right)-2m_{e}^{2} +\frac{1}{\zeta}\frac{m_{P}}{m_{e}}\frac{\varphi_{-}^{2}}{4}\left(\frac{d\bm{\mathcal{A}}_{\perp}}{d\varphi_{+}}\right)^{2}+\zeta m_{e}^{2}\left(\frac{l^{2}-m_{P}^{2}-m_{e}^{2}}{m_{e}m_{P}}\right)+\bm{x}_{\perp}^{2}\frac{m_{P}}{m_{e}}\frac{1}{\zeta}.
\end{align}
Now using well-known results on Gaussian integrals, we obtain
\begin{align}
\int l\left(\pi+\pi^{\prime}\right)e^{ig\varphi_{-}\bm{x}_{\bot}^{2}}d^{2}\bm{x}_{\perp} & =\frac{i\pi}{\varphi_{-}g}\left(\frac{m_{e}m_{P}}{\zeta}\left(1+\zeta^{2}\right)-2m_{e}^{2}+\frac{1}{\zeta}\frac{m_{P}}{m_{e}}\frac{\varphi_{-}^{2}}{4}\left(\frac{d\bm{\mathcal{A}}_{\perp}}{d\varphi_{+}}\right)^{2}+\zeta m_{e}^{2}\left(\frac{l^{2}-m_{P}^{2}-m_{e}^{2}}{m_{e}m_{P}}\right)\right)\nonumber \\
 & -\frac{\pi}{\varphi_{-}^{2}g^{2}}\frac{m_{P}}{m_{e}}\frac{1}{\zeta}
\end{align}
and therefore

\begin{align}
\int l\left(\pi+\pi^{\prime}\right)e^{i\left(\tilde{\Phi}+g\varphi_{-}\bm{x}_{\bot}^{2}\right)}d^{2}\bm{x}_{\perp}d\varphi_{-} & =\frac{if_{-1}\pi}{g}\left(\frac{m_{e}m_{P}}{\zeta}\left(1+\zeta^{2}\right)-2m_{e}^{2}+\zeta m_{e}^{2}\left(\frac{l^{2}-m_{P}^{2}-m_{e}^{2}}{m_{e}m_{P}}\right)\right)\nonumber \\
 & +\frac{\pi}{g}\frac{1}{\zeta}\frac{m_{P}}{m_{e}}c^{2}\frac{if_{1}}{4}\left(\frac{d\bm{\mathcal{A}}_{\perp}}{d\varphi_{+}}\right)^{2}-\frac{\pi}{g^{2}}c^{-1}f_{-2}\frac{m_{P}}{m_{e}}\frac{1}{\zeta}\nonumber \\
 & \approx if_{-1}2\pi(kP)m_{e}^{2}\left(1+\zeta^{2}\right)+2\pi(kP)m_{e}^{2}y^{2}if_{1}-4\pi(kP)\zeta m_{e}^{2}\frac{(kP)\left|\frac{d\bm{\mathcal{A}}_{\perp}}{d\varphi_{+}}\right|}{2m_{e}^{2}m_{P}y}f_{-2}\nonumber \\
 & =if_{-1}2\pi(kP)m_{e}^{2}\left(1+\zeta^{2}\right)+2\pi(kP)m_{e}^{2}y^{2}if_{1}-4\pi(kP)\zeta m_{e}^{2}\frac{\chi_{P}}{2y}f_{-2}.
\end{align}

\subsubsection{Term $\epsilon_{\alpha\beta\rho\sigma}l^{\alpha}\pi^{\beta}k^{\rho}\pi^{\prime\sigma}$}

By using that repeated four vectors contracted with the Levi-Civita symbol vanishes, we obtain
\begin{align}
 & \epsilon_{\alpha\beta\rho\sigma}l^{\alpha}\pi^{\beta}k^{\rho}\pi^{'\sigma}\nonumber \\
 & =\epsilon_{\alpha\beta\rho\sigma}\left(P^{\alpha}-p^{\alpha}\right)\left(p^{\beta}-A^{\beta}\right)k^{\rho}\left(p^{\sigma}-A^{\prime\sigma}\right)\nonumber \\
 & =\epsilon_{\alpha\beta\rho\sigma}P^{\alpha}\left(A^{\sigma}-A^{\prime\sigma}\right)p^{\beta}k^{\rho}\nonumber \\
 & \approx\varphi_{-}\epsilon_{\alpha\beta\rho\sigma}P^{\alpha}\frac{dA^{\sigma}}{d\varphi_{+}}p^{\beta}k^{\rho}
\end{align}
Now use the identity
\begin{equation}
g^{\mu\nu}=n^{\mu}\tilde{n}^{\nu}+n^{\nu}\tilde{n}^{\mu}-e_{1}^{\mu}e_{1}^{\nu}-e_{2}^{\mu}e_{2}^{\nu}
\end{equation}
with
\begin{align}
\tilde{n}&=\frac{1}{2}\{1,-\boldsymbol{n}\},\\
e_{i}&=\{0,\boldsymbol{e}_{i}\}
\end{align}
where $\boldsymbol{e}_{i}$ are unit vectors perpendicular to each other and to $\boldsymbol{n}$. In the setup, we have chosen, the vector potential of the laser is along $\boldsymbol{e}_{1}$ and we have
\begin{align}
 & \varphi_{-}\epsilon_{\alpha\beta\rho\sigma}P^{\alpha}\frac{dA^{\sigma}}{d\varphi_{+}}p^{\beta}k^{\rho}=\varphi_{-}\epsilon_{\alpha\beta\rho\sigma}\left[(nP)\tilde{n}^{\alpha}+(\boldsymbol{P}_{\perp}\boldsymbol{e}_{2})a_{2}^{\alpha}\right]\frac{d\mathcal{A}^{\sigma}}{d\varphi_{+}}\left[(np)\tilde{n}^{\beta}+(\boldsymbol{p}_{\perp}\boldsymbol{e}_{2})a_{2}^{\beta}\right]k^{\rho}\nonumber \\
 & =\varphi_{-}\epsilon_{\alpha\beta\rho\sigma}\left[(nP)(\boldsymbol{p}_{\perp}\boldsymbol{e}_{2})-(\boldsymbol{P}_{\perp}\boldsymbol{e}_{2})(np)\right]\frac{d\mathcal{A}^{\sigma}}{d\varphi_{+}}\tilde{n}^{\alpha}a_{2}^{\beta}k^{\rho} =\varphi_{-}\epsilon_{\alpha\beta\rho\sigma}\left[(nP)(\boldsymbol{x}_{\perp}\boldsymbol{e}_{2})\right]\frac{d\mathcal{A}^{\sigma}}{d\varphi_{+}}\tilde{n}^{\alpha}a_{2}^{\beta}k^{\rho},
\end{align}
where in the last line we put in $\bm{p}_{\perp}=\bm{x}_{\perp}+\frac{kp}{kP}\bm{P}_{\perp}+\frac{k(P-p)}{kP}\left\langle \bm{\mathcal{A}_\perp}\right\rangle $, from the change of variable, which cancels the $\boldsymbol{P}_{\perp}\boldsymbol{e}_{2}$ term. The remaining terms are linear in $\boldsymbol{x}_{\perp}$ and the integral over $\boldsymbol{x}_{\perp}$ of these terms vanishes.

\subsubsection{The total contribution from ``Line 1''}

By observing that the functions $f_{n}$ are all dimensionless and of the same order of magnitude, we may neglect terms suppressed by factors of $m_{e}/m_{P}$ or $m_{e}/m_{N}$ to obtain
\begin{equation}
\int d^{2}\bm{x}_{\perp}d\varphi_{-}T_{1}^{\mu\nu}W_{\mu\nu\alpha}l^{\alpha}e^{i\left(\tilde{\Phi}+g\varphi_{-}\bm{x}_{\bot}^{2}\right)}\approx4m_{N}m_{P}\left(g_{v}^{2}-g_{a}^{2}\right)2\pi(kP)8\left[m_{e}^{2}y^{2}if_{1}-if_{-1}m_{e}^{2}\left(1+\zeta^{2}\right)+\zeta m_{e}^{2}\frac{\chi_{P}}{y}f_{-2}\right].
\end{equation}
It is convenient to change variable from $\rho$ to a variable $z$
which should be on the order of unity. This is achieved from looking
at the definition of the variable $y$ and recognizing that the relevant
size of $\rho$ is when the term containing $\rho$ is of the same
size as the other terms. We may write
\begin{align}
y & =\sqrt{\frac{2\rho kP+m_{N}^{2}-m_{e}^{2}-m_{P}^{2}}{m_{e}m_{P}}\zeta+1+\zeta^{2}},\nonumber \\
 & =\sqrt{\frac{2\rho kP}{m_{e}m_{P}}\zeta+\frac{m_{N}^{2}-m_{e}^{2}-m_{P}^{2}}{m_{e}m_{P}}\zeta+1+\zeta^{2}}\nonumber \\
 & =\sqrt{\frac{2kP}{m_{e}m_{P}}\zeta\left[\rho+\frac{m_{N}^{2}-m_{e}^{2}-m_{P}^{2}}{2\zeta kP}\zeta+\frac{m_{e}m_{P}}{2\zeta kP}\left(1+\zeta^{2}\right)\right]},
\end{align}
and therefore we introduce
\begin{align}
z & =\frac{\rho}{\frac{m_{N}^{2}-m_{e}^{2}-m_{P}^{2}}{2\zeta kP}\zeta+\frac{m_{e}m_{P}}{2\zeta kP}\left(1+\zeta^{2}\right)}=\zeta\frac{2kP\rho}{\left(m_{N}^{2}-m_{e}^{2}-m_{P}^{2}\right)\zeta+m_{e}m_{P}\left(1+\zeta^{2}\right)}.
\end{align}
In this way, the contribution to the probability is
\begin{align}
dP & =\frac{G_{F}^{2}}{2}\frac{1}{(2\pi)^{6}}\frac{1}{32kP}\int d\rho d\varphi_{+}\frac{J}{2l^{2}}(l^{2}-m_{N}^{2})\int T_{1}^{\mu\nu}W_{\mu\nu\alpha}l^{\alpha}e^{i\left(\tilde{\Phi}+g\varphi_{-}x_{\bot}^{2}\right)}d^{2}\bm{x}_{\perp}d\varphi_{-}\frac{dp_{\parallel}}{\varepsilon_{p}}
\end{align}
and changing variable from $\rho$ to $z$, writing $d\varphi_{+}=kudt=kUd\tau=\frac{kP}{m_{P}}d\tau$,
$dp_{\parallel}/\varepsilon_{p}=d(kp)/kp=d\zeta/\zeta$ and
\begin{equation}
\frac{J}{2l^{2}}(l^{2}-m_{N}^{2})=\frac{2\pi}{2l^{4}}(l^{2}-m_{N}^{2})^{2}=\frac{2\pi}{2l^{4}}z^{2}\left[\frac{\left(m_{N}^{2}-m_{e}^{2}-m_{P}^{2}\right)\zeta+m_{e}m_{P}\left(1+\zeta^{2}\right)}{\zeta}\right]^{2},
\end{equation}
we obtain
\begin{align}
dP & =\frac{G_{F}^{2}}{2}\frac{1}{(2\pi)^{6}}\frac{1}{32kP}\int\frac{kP}{m_{P}}d\tau dz\frac{z^{2}}{2kP}\left[\left(m_{N}^{2}-m_{e}^{2}-m_{P}^{2}\right)\zeta+m_{e}m_{P}\left(1+\zeta^{2}\right)\right]^{3}\frac{d\zeta}{\zeta^{4}}\nonumber \\
 & \times\frac{2\pi}{2l^{4}}4m_{N}m_{P}\left(g_{v}^{2}-g_{a}^{2}\right)2\pi(kP)8\left[m_{e}^{2}y^{2}if_{1}-if_{-1}m_{e}^{2}\left(1+\zeta^{2}\right)+\zeta m_{e}^{2}\frac{\chi_{P}}{y}f_{-2}\right]\nonumber \\
\nonumber \\
 & =\frac{G_{F}^{2}}{2}\frac{1}{(2\pi)^{4}}m_{e}^{5}m_{P}^{2}\int dzd\tau d\zeta\frac{z^{2}}{\zeta^{4}}\frac{1}{8l^{4}}\left[\frac{\left(m_{N}^{2}-m_{e}^{2}-m_{P}^{2}\right)}{m_{e}m_{P}}\zeta+\left(1+\zeta^{2}\right)\right]^{3}\nonumber \\
 & \times2m_{N}m_{P}\left(g_{v}^{2}-g_{a}^{2}\right)\left[y^{2}if_{1}-if_{-1}\left(1+\zeta^{2}\right)+\zeta\frac{\chi_{P}}{y}f_{-2}\right].
\end{align}

\subsection{Line 2 (symmetric part)}

We start by using the identity
\begin{equation}
l_{\alpha}l^{\beta}T_{2}^{\mu\nu\alpha}W_{\mu\nu\beta}=l_{\alpha}l^{\beta}T_{2,S}^{\mu\nu\alpha}W_{\mu\nu\beta}^{S}+l_{\alpha}l^{\beta}T_{2,A}^{\mu\nu\alpha}W_{\mu\nu\beta}^{A}.
\end{equation}
Here, we evaluate
\begin{equation}
\int d^{2}\bm{x}_{\perp}d\varphi_{-}l_{\alpha}l^{\beta}T_{2,S}^{\mu\nu\alpha}W_{\mu\nu\beta}^{S}e^{i\left(\tilde{\Phi}+g\varphi_{-}\bm{x}_{\bot}^{2}\right)}.
\end{equation}
where the labels $S$ and $A$ indicate the symmetric and the anti-symmetric parts of the corresponding tensors. After performing straightforward manipulations we obtain
\begin{align}
 & l_{\alpha}l^{\beta}T_{2,S}^{\mu\nu\alpha}W_{\mu\nu\beta}^{S}\nonumber \\
 &=16\left\{\left(g_{v}^{2}+g_{a}^{2}\right)\left[ \frac{lk}{kP}\left(\Pi\Pi^{\prime}-m_{P}^{2}\right)-l\left(\Pi+\Pi^{\prime}\right)\right] -\frac{2g_{v}g_{a}}{kP}i\epsilon^{\alpha\beta\rho\sigma}l_{\alpha}\Pi_{\beta}k_{\rho}\Pi_{\sigma}^{\prime}\right\}\nonumber \\
 & \times\left\{\left[ \frac{lk}{kp}\left(\pi\pi^{\prime}-m_{e}^{2}\right)-l\left(\pi+\pi^{\prime}\right)\right] +\frac{1}{kp}i\epsilon_{\tau\theta\phi\kappa}l^{\tau}\pi^{\theta}k^{\phi}\pi^{\prime\kappa}\right\}\nonumber \\
 & +16l^{2}\left(g_{v}^{2}+g_{a}^{2}\right)\left[\left(\Pi^{\prime}+\Pi\right)\cdot\left(\pi^{\prime}+\pi\right)-2kP\frac{\pi\pi^{\prime}-m_{e}^{2}}{kp}-2kp\frac{\Pi\Pi^{\prime}-m_{P}^{2}}{kP}\right]\nonumber \\
 & +16l^{2}\left[\frac{2g_{v}g_{a}}{kP}i\epsilon^{\mu\beta\rho\sigma}\left(\pi_{\mu}^{\prime}+\pi_{\mu}\right)\Pi_{\beta}k_{\rho}\Pi_{\sigma}^{\prime}-\frac{1}{kp}i\epsilon_{\mu\theta\phi\kappa}\left(\Pi^{\prime\mu}+\Pi^{\mu}\right)\pi^{\theta}k^{\phi}\pi^{\prime\kappa}\right].
\end{align}
In this expression, we encounter some terms of the same type as before and also
some new ones. The terms of the type $\epsilon^{\mu\beta\rho\sigma}\left(\pi_{\mu}^{\prime}+\pi_{\mu}\right)\Pi_{\beta}k_{\rho}\Pi_{\sigma}^{\prime}$
vanish under our assumption of linear polarization.

\subsubsection{Term $l\left(\Pi+\Pi^{\prime}\right)$}

We have that
\begin{equation}
l\left(\Pi+\Pi^{\prime}\right)=(P-p+sk)\left(2P-(\mathcal{A}+\mathcal{A}^{\prime})+\frac{1}{kP}\left[\frac{\bm{\mathcal{A}}_{\perp}^{2}+\bm{\mathcal{A}}_{\perp}^{\prime2}}{2}-\bm{P}_{\perp}(\bm{\mathcal{A}}_{\perp}+\mathcal{A}{}_{\perp}^{\prime})\right]k\right),
\end{equation}
which, after performing the same kind of reduction as for $l\left(\pi+\pi^{\prime}\right)$,
leads to
\begin{align}
l\left(\Pi+\Pi^{\prime}\right) & =2m_{P}^{2}+\left(kp\frac{m_{P}^{2}}{kP}+kP\frac{m_{e}^{2}}{kp}\right)\frac{kp}{k\left(P-p\right)}+\frac{k(P-p)}{kP}\frac{(\bm{\mathcal{A}}_{\perp}-\bm{\mathcal{A}}_{\perp}^{\prime})^{2}}{4}+2\rho kP\nonumber \\
 & +kP\frac{m_{N}^{2}-m_{P}^{2}-m_{e}^{2}}{k\left(P-p\right)}+\bm{x}_{\perp}^{2}\frac{kP}{k\left(P-p\right)}.
\end{align}

\subsubsection{The total contribution from the symmetric part of ``Line 2''}

We note that several terms vanish when performing the integration over $d^{2}\bm{x}_{\perp}$ in this contribution. We have
\begin{align}
 & \int d^{2}\bm{x}_{\perp}l_{\alpha}l^{\beta}T_{2,S}^{\mu\nu\alpha}W_{\mu\nu\beta}^{S}\nonumber \\
 & =\int d^{2}\bm{x}_{\perp}16\left(g_{v}^{2}+g_{a}^{2}\right)\left[ \frac{lk}{kP}\left(\Pi\Pi^{\prime}-m_{P}^{2}\right)-l\left(\Pi+\Pi^{\prime}\right)\right] \left[ \frac{lk}{kp}\left(\pi\pi^{\prime}-m_{e}^{2}\right)-l\left(\pi+\pi^{\prime}\right)\right] \nonumber \\
 & +32\frac{kP}{kp}g_{v}g_{a}(\bm{x}_{\perp}\boldsymbol{e}_{2})^{2}\left(\bm{\mathcal{A}}_{\perp}-\bm{\mathcal{A}}_{\perp}^{\prime}\right)^{2}\nonumber \\
 & +16l^{2}\left(g_{v}^{2}+g_{a}^{2}\right)\left[\left(\Pi^{\prime}+\Pi\right)\left(\pi^{\prime}+\pi\right)-2kP\frac{\pi\pi^{\prime}-m_{e}^{2}}{kp}-2kp\frac{\Pi\Pi^{\prime}-m_{P}^{2}}{kP}\right].
\end{align}
Now using that in the LCFA and to leading order in $m_{e}/m_{P}$
\begin{align}
l\left(\pi+\pi^{\prime}\right)&\approx\left(kp\frac{m_{P}^{2}}{kP}+kP\frac{m_{e}^{2}}{kp}\right)+\frac{kP}{kp}\frac{\varphi^{2}}{4}\left(\frac{d\bm{\mathcal{A}}_{\perp}}{d\varphi_+}\right)^{2}+(l^{2}-m_{N}^{2})\frac{kp}{kP}+\bm{x}_{\perp}^{2}\frac{kP}{kp},\\
l\left(\Pi+\Pi^{\prime}\right)&\approx2m_{P}^{2}+\frac{\varphi^{2}}{4}\left(\frac{d\bm{\mathcal{A}}_{\perp}}{d\varphi_+}\right)^{2}+(l^{2}-m_{N}^{2})+\bm{x}_{\perp}^{2},\\
\left(\Pi+\Pi^{\prime}\right)\left(\pi+\pi^{\prime}\right) & \approx2kP\frac{m_{e}^{2}+\bm{x}_{\perp}^{2}}{kp}+2kp\frac{m_{P}^{2}}{kP}+2\left(\frac{kP}{kp}+\frac{kp}{kP}\right)\frac{(\bm{\mathcal{A}}_{\perp}-\bm{\mathcal{A}}^{\prime}_{\perp})^{2}}{4}
\end{align}
we obtain that to leading order
\begin{align}
 & \int d^{2}\bm{x}_{\perp}d\varphi_{-}l_{\alpha}l^{\beta}T_{2,S}^{\mu\nu\alpha}W_{\mu\nu\beta}^{S}\nonumber \\
 & =16\left(g_{v}^{2}+g_{a}^{2}\right)2\pi(kP)\left(2m_{P}^{2}+3l^{2}-m_{N}^{2}\right)\nonumber \\
 & \times\left\{-m_{e}^{2}y^{2}if_{1}-m_{e}^{2}\zeta\frac{\chi_{P}}{y}f_{-2}+m_{e}^{2}if_{-1}\left[1+\left( 1+\frac{2m_{P}^{2}+l^{2}-m_{N}^{2}}{2m_{P}^{2}+3l^{2}-m_{N}^{2}}\frac{l^{2}-m_{N}^{2}}{m_{P}^{2}}\right) \zeta^{2}\right]\right\}
\end{align}

\subsection{Line 2 (anti-symmetric part)}

By proceeding analogously as above, we obtain
\begin{align}
 & l_{\alpha}l^{\tau}T_{2,A}^{\mu\nu\alpha}W_{\mu\nu\tau}^{A}\nonumber \\
 & =\frac{16\left(g_{v}^{2}+g_{a}^{2}\right)}{(kP)(kp)}\left\{(lk)^{2}\left[ \left(\Pi^{\prime}\pi^{\prime}\right)\left(\Pi\pi\right)-\left(\Pi^{\prime}\pi\right)\left(\Pi\pi^{\prime}\right)\right] \right.\nonumber \\
 & +(lk)(kP)\left[(l\pi)\pi^{\prime}-(l\pi^{\prime})\pi\right]\cdot\left(\Pi-\Pi^{\prime}\right)\nonumber \\
 & +(lk)(kp)\left[(l\Pi)\Pi^{\prime}-(l\Pi^{\prime})\Pi\right]\cdot\left(\pi-\pi^{\prime}\right)\nonumber \\
 & \left.-(kp)(kP)\left(l\Pi-l\Pi^{\prime}\right)\left(l\pi-l\pi^{\prime}\right)\right\}\nonumber \\
 & -\frac{16\left(g_{v}^{2}+g_{a}^{2}\right)}{kP}\left\{ \left[\left(lk\right)\Pi^{\prime\nu}\Pi^{\mu}+\left(l\Pi\right)k^{\nu}\Pi^{\prime\mu}+\left(l\Pi^{\prime}\right)k^{\mu}\Pi^{\nu}\right]\left(\pi^{\beta}+\pi^{\prime\beta}\right)+\frac{lk}{kp}\Pi^{\prime\nu}\Pi^{\mu}\left(m_{e}^{2}-\pi\pi^{\prime}\right)k^{\beta}\right\} i\epsilon_{\mu\nu\alpha\beta}l^{\alpha}\nonumber \\
 & +\frac{32g_{v}g_{a}}{kp}\left\{ \left[\left(lk\right)\pi_{\nu}^{\prime}\pi_{\mu}+\left(l\pi\right)k_{\nu}\pi_{\mu}^{\prime}+\left(l\pi^{\prime}\right)k_{\mu}\pi_{\nu}\right]\left(\Pi_{\beta}+\Pi_{\beta}^{\prime}\right)+\frac{lk}{kP}\pi_{\nu}^{\prime}\pi_{\mu}\left(m_{P}^{2}-\Pi\Pi^{\prime}\right)k_{\beta}\right\} i\epsilon^{\mu\nu\alpha\beta}l_{\alpha}\nonumber \\
 & +\frac{32g_{v}g_{a}}{(kP)(kp)}\left\{l^{2}\left[ (kP)(kp)\left(\pi+\pi^{\prime}\right)\cdot\left(\Pi+\Pi^{\prime}\right)+2(kP)^{2}\left(m_{e}^{2}-\pi\pi^{\prime}\right)+2(kp)^{2}\left(m_{P}^{2}-\Pi\Pi^{\prime}\right)\right] \right.\nonumber \\
 & \left.-\left[ (kP)l\cdot\left(\Pi+\Pi^{\prime}\right)+(lk)\left(m_{P}^{2}-\Pi\Pi^{\prime}\right)\right] \left[ (kp)l\cdot\left(\pi+\pi^{\prime}\right)+(lk)\left(m_{e}^{2}-\pi\pi^{\prime}\right)\right] \right\}
\end{align}
This whole part, however, only contributes with terms suppressed by at least $m_{e}/m_{P}$.

\subsection{Line 3}

After performing reductions from the initial expression, one obtains
\begin{align}
T_{2,S}^{\mu\nu\alpha}W_{\mu\nu\alpha}^{S} & =80\left(g_{v}^{2}+g_{a}^{2}\right)\left[\left(\Pi^{\prime}+\Pi\right)\cdot\left(\pi^{\prime}+\pi\right)-2kp\frac{\Pi\Pi^{\prime}-m_{P}^{2}}{kP}-2kP\frac{\pi\pi^{\prime}-m_{e}^{2}}{kp}\right]\nonumber \\
 & +\frac{160g_{v}g_{a}}{kP}i\epsilon^{\mu\beta\rho\sigma}\left(\pi_{\mu}^{\prime}+\pi_{\mu}\right)\Pi_{\beta}k_{\rho}\Pi_{\sigma}^{\prime}-\frac{80\left(g_{v}^{2}+g_{a}^{2}\right)}{kp}i\epsilon_{\mu\theta\phi\kappa}\left(\Pi^{\prime\mu}+\Pi^{\mu}\right)\pi^{\theta}k^{\phi}\pi^{\prime\kappa}\nonumber \\
 & +160g_{v}g_{a}\left(\Pi^{\prime}-\Pi\right)\left(\pi'-\pi\right).
\end{align}
By using the identity
\begin{equation}
\left(\Pi-\Pi^{\prime}\right)\cdot\left(\pi^{\prime}-\pi\right)=\left(\bm{\mathcal{A}}^{\prime}_{\perp}-\bm{\mathcal{A}}_{\perp}\right)^{2},
\end{equation}
it is seen that this contribution contains only terms which have already been analyzed.
For the anti-symmetric part, we obtain after reduction
\begin{align}
T_{2,A}^{\mu\nu\alpha}W_{\mu\nu\alpha}^{A} & =48\left(g_{v}^{2}+g_{a}^{2}\right)\left(\Pi-\Pi^{\prime}\right)\cdot\left(\pi^{\prime}-\pi\right)\nonumber \\
 & -\frac{48\left(g_{v}^{2}+g_{a}^{2}\right)}{kP}i\epsilon_{\mu\nu\alpha\beta}\left(\pi^{\beta}+\pi^{\prime\beta}\right)k^{\alpha}\Pi^{\prime\nu}\Pi^{\mu}\nonumber \\
 & +\frac{96g_{v}g_{a}}{kp}i\epsilon^{\mu\nu\alpha\beta}\left(\Pi_{\beta}+\Pi_{\beta}^{\prime}\right)k_{\alpha}\pi_{\nu}^{\prime}\pi_{\mu}\nonumber \\
 & -96g_{v}g_{a}\left[\left(\Pi+\Pi^{\prime}\right)\cdot\left(\pi+\pi^{\prime}\right)+\frac{2kP}{kp}\left(m_{e}^{2}-\pi\pi^{\prime}\right)+\frac{2kp}{kP}\left(m_{P}^{2}-\Pi\Pi^{\prime}\right)\right],
\end{align}
which also contains only terms which we have already treated.

\end{widetext}


%

\end{document}